\documentclass[12pt]{article}
\usepackage{epsf, cite, amssymb}
\usepackage{epsfig}
\usepackage{amsmath}
\usepackage[matrix,arrow,frame,import,curve,color]{xy}
\usepackage[bookmarks=true]{hyperref}
\usepackage[active]{srcltx}

\makeatletter
\renewcommand\section{\@startsection {section}{1}{\z@}%
                                   {-3.5ex \@plus -1ex \@minus -.2ex}%nn
                                   {2.3ex \@plus.2ex}%
                                   {\normalfont\large\bfseries}}
\renewcommand\subsection{\@startsection{subsection}{2}{\z@}%
                                     {-3.25ex\@plus -1ex \@minus -.2ex}%
                                     {1.5ex \@plus .2ex}%
                                     {\normalfont\bfseries}}
\makeatother

\newcommand{\nc}{\newcommand}
\def\cDb{\overline{\cD}}
\def\cQb{\overline{\cQ}}

\def\Aut{\operatorname{Aut}}
\def\Autt{\widetilde{\Aut}}

 \nc{\ra}{\rightarrow}
\def\al{\alpha}

\nc{\ve}{\varepsilon}
\def\gam{\gamma}

\def\om{\omega}
\def\tha{\theta}

\def\Gam{\Gamma}

\def\Om{\Omega}
\def\Sig{\Sigma}

\nc{\bea}{\begin{eqnarray}}
\nc{\eea}{\end{eqnarray}}
\nc{\be}{\begin{equation}}
\nc{\ee}{\end{equation}}
\nc{\cA}{{\cal A}}
\nc{\cB}{ \cal B}

\def\cD{{\cal D}}

\nc{\cF}{{\cal F}}
\nc{\cG}{{\cal G}}
\nc{\cg}{{\cal g}}

\def\cJ{{\cal J}}

\def\cK{{\cal K}}
\nc{\cL}{{\cal L}}
\nc{\M}{{\cal M}}
\nc{\cM}{{\cal M}}
\def\N{{\cal N}}

\def\cO{{\cal O}}

\nc{\cQ}{{\cal Q}}
\nc{\cR}{{\cal R}}

\def\T{{\cal T}}

\nc{\BB}{{\mathbb B}}
\nc{\CC}{{\mathbb C}}

\nc{\DD}{{\mathbb D}}
\nc{\EE}{{\mathbb E}}
\nc{\FF}{{\mathbb F}}
\nc{\GG}{{\mathbb G}}
\nc{\HH}{{\mathbb H}}
\nc{\JJ}{{\mathbb J}}
\nc{\RR}{{\mathbb R}}
\nc{\R}{{\mathbb R}}
\nc{\MM}{{\mathbb M}}
\nc{\PP}{{\mathbb P}}
\def\P{{\mathbb P}}
\nc{\QQ}{{\mathbb Q}}
\nc{\ZZ}{{\mathbb Z}}
\nc{\CP}{{\CC\PP}}
\nc{\calone}{{\mathbb 1}}
\nc{\half}{\frac{1}{2}}
\nc{\qrt}{\frac{1}{4}}
\nc{\del}{\partial}

\nc{\delbar}{\bar\partial}
\nc{\Spin}{{\rm Spin}}
\nc{\SO}{\operatorname{SO}}
\nc{\CS}{{\rm CS}}
\renewcommand{\O}{{\mathcal O}}
\nc{\Sp}{{\rm Sp}}

\nc{\com}[2]{{ \left[ #1, #2 \right] }}
\nc{\acom}[2]{{ \left\{ #1, #2 \right\} }}
\nc{\rr}{\rightarrow}
\nc{\p}{\partial}
\nc{\LT}{{\LL_\T}}
\nc{\Tr}{{\rm Tr}}
\nc{\tr}{{\rm tr}}
\nc{\ch}{{\rm ch}}
\def\com#1#2{{ \left[ #1, #2 \right] }}
\def\acom#1#2{{ \left\{ #1, #2 \right\} }}
\nc{\wt}{\widetilde}
\nc{\tKT}{\widetilde{K3}}
\nc{\ttha}{\tilde{\theta}}
\nc{\tphi}{\tilde{\phi}}
\nc{\tPhi}{\tilde{\Phi}}
\nc{\tpsi}{\tilde{\psi}}
\nc{\tgam}{\tilde{\gam}}
\nc{\tGam}{\tilde{\Gam}}
\nc{\tSig}{\tilde{\Sig}}
\nc{\tc}{\tilde c}

\nc{\te}{\tilde e}
\nc{\tg}{\tilde g}
\nc{\tj}{\tilde j}
\nc{\tp}{\widetilde{p}}
\nc{\tq}{\widetilde{q}}
\nc{\ts}{{\tilde s}}
\nc{\tz}{\tilde z}
\nc{\tA}{{\tilde A}}
\nc{\tD}{{\tilde D}}
\nc{\tE}{{\tilde E}}
\nc{\tG}{{\tilde G}}
\nc{\tH}{{\tilde H}}
\nc{\tM}{{\tilde M}}
\nc{\tN}{{\tilde N}}
\nc{\tP}{{\tilde P}}
\nc{\tQ}{{\tilde Q}}
\nc{\tS}{\tilde{S}}
\nc{\tF}{\tilde{{\cal F}}}
\nc{\tX}{\widetilde{X}}
\def\ff#1#2{{\textstyle\frac{#1}{#2}}}

\nc{\wh}{\widehat}
\nc{\hd}{\hat d}
\nc{\he}{\hat e}
\nc{\hf}{\hat f}
\nc{\hg}{\hat g}
\nc{\hh}{\hat h}
\nc{\hp}{\hat p}
\nc{\hw}{\hat w}
\nc{\hx}{\hat x}
\nc{\hy}{\hat y}
\nc{\hz}{\hat z}
\nc{\hA}{\widehat{A}}
\nc{\hE}{\widehat{E}}
\nc{\hH}{\widehat{H}}
\nc{\hJ}{\widehat{J}}
\nc{\tK}{\widetilde{K}}
\nc{\hM}{\widehat M}
\nc{\hF}{\widehat{\F}}

\nc{\ha}{\widehat \alpha}
\nc{\hphi}{\hat{\phi}}
\nc{\hpsi}{\hat{\psi}}
\nc{\hgam}{\hat{\gam}}
\nc{\hPhi}{\hat{\Phi}}
\nc{\hPsi}{\hat{\Psi}}
\nc{\hGam}{\hat{\Gam}}

\nc{\w}{\wedge}

\nc{\ol}{\overline}
\nc{\abar}{{\ol{a}}}
\nc{\bbar}{{\ol{b}}}
\nc{\cbar}{{\ol{c}}}
\nc{\ebar}{{\ol{e}}}
\nc{\ibar}{{\ol{\imath}}}
\nc{\jbar}{{\ol{\jmath}}}
\nc{\kbar}{{\ol{k}}}
\nc{\lbar}{{\ol{l}}}
\nc{\mbar}{{\ol{m}}}
\nc{\nbar}{{\ol{n}}}
\nc{\ubar}{{\ol{u}}}
\nc{\vbar}{{\ol{v}}}
\nc{\wbar}{{\ol{w}}}
\nc{\xbar}{{\ol{x}}}
\nc{\ybar}{{\ol{y}}}
\nc{\zbar}{{\ol{z}}}

\nc{\Xbar}{{\overline X}}
\nc{\Zbar}{{\overline Z}}

\nc{\epsbar}{\ol{\epsilon}}
\nc{\lambar}{\ol{\lambda}}
\nc{\psibar}{\ol{\psi}}
\nc{\Psibar}{\ol{\Psi}}
\nc{\phibar}{\ol{\phi}}

\nc{\Phibar}{\ol{\Phi}}
\nc{\chibar}{\ol{\chi}}
\nc{\ombar}{\ol{\om}}
\nc{\Ombar}{\ol{\Om}}

\nc{\alphabar}{{\ol{\alpha}}}
\nc{\betabar}{{\ol{\beta}}}
\nc{\tsevbar}{\ol{\bf 27}}
\nc{\tsev}{{\bf 27}}
\nc{\bah}{{\mathbf {\hat{A}}}}
\nc{\bX}{{\mathbf X}}

\nc{\dal}{\dot{\al}}
\nc{\thab}{\bar{\theta}}
\nc{\thal}{\theta^{\al}}
\nc{\thdal}{\bar{\theta}^{\dal}}

\nc{\thsigthm}{\tha \sigma^m \thab}
\nc{\thsigthn}{\tha \sigma^n \thab}

\nc{\Dal}{D_{\al}}
\nc{\Ddal}{\bar{D}_{\dal}}
\nc{\CDal}{{\cal D}_{\al}}
\nc{\CDdal}{\bar{\cal D}_{\dal}}

\nc{\eq}[1]{(\ref{#1})}
\nc{\non}{\nonumber}
\nc{\comment}[1]{\noindent{\bf #1}}
\nc{\fcomment}[1]{\footnote{{\bf #1}}}
\nc{\bwcomment}[1]{\footnote{{\bf[\textcolor{red}{ #1}}]}}
\nc{\xs}{\not\!\!X}
\nc{\ps}{\not\!\!P}
\nc{\dif}{{d}}
\nc{\equ}{{\rm eq}}

\nc{\AdS}{{\rm AdS}}
\nc{\vol}{{\rm vol}}
\nc{\Ainf}{A_{\infty}}
\nc{\End}{{\rm End}}
\nc{\Ext}{{\rm Ext}}
\nc{\Hom}{{\rm Hom}}
\nc{\IIB}{{\rm IIB}}
\nc{\Pic}{{\rm Pic}}
\nc{\SL}{{\rm SL}}
\nc{\GL}{{\rm GL}}
\nc{\Ker}{{\rm Ker}}
\nc{\diag}{{\rm diag}}

\nc{\bra}[1]{\langle{#1}|}
\nc{\ket}[1]{|{#1}\rangle}
\nc{\braket}[2]{\langle{#1}|{#2}\rangle}
\nc{\sect}[1]{Section~\ref{#1}}
\nc{\fig}[1]{Fig.~\ref{#1}}
\nc{\chap}[1]{Chapter~\ref{#1}}
\nc{\Dslash}{\ensuremath \raisebox{0.025cm}{\slash}\hspace{-0.32cm} D}
\nc{\Hslash}{\ensuremath \raisebox{0.025cm}{\slash}\hspace{-0.32cm} H}
\nc{\no}{\!:\!\!}
\nc{\bpm}{\begin{pmatrix}}
\nc{\epm}{\end{pmatrix}}
 \nc{\bi}{\begin{itemize}}
 \nc{\ei}{\end{itemize}}
 \nc{\ben}{\begin{enumerate}}
 \nc{\een}{\end{enumerate}}

\newcommand{\C}[1]{$(\ref{#1})$}

\def\Z{\mathbb{Z}}

\def\ff#1#2{{\textstyle\frac{#1}{#2}}}
\def\SO{\operatorname{SO}}

\def\SU{\operatorname{SU}}
\def\U{\operatorname{U}}
\def\GU{\operatorname{U{}}}

\nc{\rank}{{\rm rank}}
\nc{\pr}{{\rm pr}}

\nc{\tom}{\tilde{\om}}
\nc{\tOm}{\tilde{\Om}}

\def\M{\mathcal{M}}

\def\bz{{\ol{z}}}

\def\Up{{\Upsilon}}

\newcommand\gammah{\widehat{\gamma}}
\newcommand\deltah{\widehat{\delta}}

\newcommand\kappah{\widehat{\kappa}}

\newcommand\sigmah{\widehat{\sigma}}

\newcommand\gammab{\overline{\gamma}}

\newcommand\thetab{\overline{\theta}}

\newcommand\Gammab{\overline{\Gamma}}

\newcommand\Sigmab{\overline{\Sigma}}
\newcommand\Upsilonb{\overline{\Upsilon}}
\newcommand\Phib{\overline{\Phi}}

\newcommand\ah{\widehat{a}}

\newcommand\zb{\overline{z}}

\newcommand\Qh{\widehat{Q}}

\newcommand\Eb{\overline{E}}

\newcommand\Jt{\widetilde{J}}

\def\la{\langle}
\def\ra{\rangle}
\def\lad{\langle\!\langle}
\def\rad{\rangle\!\rangle}

\def\rep#1{{{\boldsymbol{#1}}}}
\def\brep#1{{{\overline{\boldsymbol{#1}}}}}
\def\ri{\operatorname{ri}}
\def\bdelta{{\boldsymbol{\delta}}}
\def\bdeltah{{\boldsymbol{\deltah}}}
\def\bgamma{{\boldsymbol{\gamma}}}
\def\bgammah{{\boldsymbol{\gammah}}}

\def\bsigma{{\boldsymbol{\sigma}}}
\def\bSigma{{\boldsymbol{\Sigma}}}
\def\bE{{\boldsymbol{E}}}

\def\bsigmah{{\boldsymbol{\sigmah}}}

\begin{document}
\begin{titlepage}

\begin{center}

%{September 26, 2005}
\today \hfill         \phantom{xxx} \hfill DAMTP-2010-66

\vskip 2 cm {\Large \bf The Revival of $(0,2)$ Linear Sigma Models}\non\\
\vskip 1.25 cm { Jock McOrist\footnote{j.mcorist@damtp.cam.ac.uk}\non\\
{\vskip 0.5cm   DAMTP,
Centre for Mathematical Sciences,
 Wilberforce Road,
  Cambridge,
   CB3 OWA, UK}\non\\ }

\end{center}
\vskip 2 cm

\begin{abstract}
\baselineskip=18pt
Compactifications of the heterotic string are a viable route to phenomenologically realistic vacua and interesting new mathematics. While supergravity aspects of heterotic compactifications are largely well-understood their worldsheet description remains largely unexplored. We review recent work in developing linear sigma model techniques aimed at elucidating the underlying worldsheet description.

\end{abstract}

\end{titlepage}

\pagestyle{plain}
%\baselineskip=18pt
% Try a wider skip
\baselineskip=19pt
\newpage
\tableofcontents
\newpage

%%%%%%%%%%%%%%%%%%%%%%%%%%%%%%%%%%%%%%%%%%%%%%%%%%%%%%%%%%%%%%%%%%%%%%%%%%%%%%

\section{Introduction}
A generic compactification of string theory preserving $\N=1$ supersymmetry in $d=4$ is most likely strongly coupled, involving non-trivial fluxes and non-trivial metrics\cite{Dine:1985kv,Dine:1985he}. The most well-studied class of string compactifications arise in the type II string, generically involving Ramond-Ramond (RR) fluxes and orientifolds (see for example \cite{Grana:2005jc}). The analysis of such backgrounds is typically restricted to the supergravity approximation largely due to the difficulties in understanding RR fluxes and orientifolding from the point of view of the worldsheet (though see for example \cite{Linch:2008rw}). However, supergravity is really only a valid approximation when all the length scales in the problem are large compared to the string length. If this is not the case, then $\alpha'$ corrections become non-negligible and one needs to take them into account. As is known in other contexts, $\alpha'$ corrections can qualitatively modify the physics of the compactification. \footnote{Although \cite{Dine:1985he} argue string theory is likely to have strong coupling $g_s$ effects, we will not discuss these in this review, instead concentrating on $\alpha'$ issues relevant to \cite{Dine:1985kv}.} For example, quantum effects are known to resolve classical singularities, to connect seemingly disparate geometric spaces, provide tests of string dualities and destabilise classical string vacua (i.e. vacua that exist in supergravity but not in string theory). In fact, compactifications defined by Landau-Ginzburg theories have no conventional notion of a target space geometry at all, and such vacua do not have a supergravity description.  Thus, in order to understand the role of quantum corrections in compactifications of perturbative string theory we really need to have a worldsheet description.

We will review some recent developments in understanding the worldsheet aspects of $\N=1$ compactifications of the heterotic string. We focus on the heterotic string for a number of reasons. Firstly, its degrees of freedom are cleaner and simpler to study on the worldsheet than their type II cousins: compactifications are constructed purely from NSNS fields, the string coupling is often tunable and there is no need to orientifold. Secondly, compactifications of the heterotic string naturally give rise to phenomenologically interesting four-dimensional vacua. For example, the existence of the gauge field means it is easy to produce chiral theories in four-dimensions and by a judicious choice of target space geometry one can construct a compactification that closely resembles the standard model. Finally, studying type II string compactifications on Calabi-Yau manifolds lead to interesting new mathematical insights (e.g. mirror symmetry). As the heterotic string target space consists of a six-dimensional geometry together with choice of gauge bundle, studying heterotic compactifications via the worldsheet will likely lead to similar insights and progress in understanding the mathematics of gauge bundles.

A broad classification of heterotic string vacua begins with the existence of a large volume limit. If the compactification has a limit where all length scales of the compactification become large compared to the string scale, then supergravity is a good description of the background, and as such may be used as the semi-classical starting point for defining a $(0,2)$-worldsheet theory. Alternatively, if one or more cycles are fixed to be string size, the compactification is much harder to define. Supergravity is not necessarily a good description of the background, and typically does not (except perhaps in special cases) give a nice semi-classical starting point for string perturbation theory. Indeed, the only known examples have been constructed via duality\cite{Dasgupta:1999ss} (see also \cite{Becker:2003yv,Becker:2003gq,Becker:2006et,Becker:2003sh,Becker:2009zx,Becker:2002sx,Becker:2009df,Becker:2005nb,Becker:2008rc}). \footnote{These spaces may in addition be characterised that their topology and complex structure do not admit K\"ahler metrics, as is argued in \cite{Goldstein:2002pg} when the geometry admits a $T^2$ fibration.} The lack of a large volume limit means constructing a well-defined worldsheet theory is difficult and our current understanding of such constructions is quite limited; though some progress has been made recently in this direction by \cite{Adams:2006kb,Adams:2009av,Adams:2009tt}. %\footnote{Dualities can also be used to understand the generic type of heterotic vacua. For example, in \cite{McOrist:2010jw}, one is able to use F-theory/Heterotic duality to argue that the generic heterotic vacuum is essentially non-geometric, not having a classical notion of geometry at all. Instead, it is intrinsically quantum mechanical.}
For this reason, we will be interested in spaces with a large volume limit, as we then stand a much better chance of understanding the role of $\alpha'$ corrections.

Compactifications with a large volume limit that preserve supersymmetry are specified by a Calabi-Yau manifold $M$ and a choice of holomorphic vector bundle $\cF$ which satisfies a Bianchi identity relating the Chern classes of $\cF$ and $M$ to the heterotic $B_2$-field.\footnote{Many good textbooks explain the heterotic Bianchi identity as well as the standard embedding solution. Two nice examples include  \cite{Green:1987mn,Polchinski:1998rr}}. An easy way to satisfy this Bianchi identity is by identifying $\cF$ with the tangent bundle. This gives rise to an unbroken $E_6$ gauge group in spacetime, with matter fields in the $\tsev$ and $\tsevbar$ representations of $E_6$. There is also a nice worldsheet description, being described by $(2,2)$ conformal field theories. However, although an easy way to satisfy the Bianchi identity, such compactifications do not give rise to a realistic phenomenology. One way to improve the situation is a more sophisticated choice of vector bundle. Then it is easy to generate spacetime GUT groups like $E_6$, $\SO(10)$ and $\SU(5)$, and there has been much progress in realising standard model and GUT like scenarios in heterotic supergravity (for example, some recent references include \cite{Andreas:2007ei,Braun:2005nv,Anderson:2007nc,Bouchard:2005ag,Buchmuller:2005jr}). The particle spectrum can be constructed using well-known methods from algebraic geometry, and Wilson lines can break the GUT group down to the standard model gauge group $\SU(3)\times\SU(2)\times \U(1)$. Do such compactifications admit worldsheet descriptions? Although in principle the answer is yes, a systematic study of such worldsheet conformal field theories is distinctly lacking in the literature. In this review we attempt to ameliorate this by systematically studying heterotic string compactifications via the worldsheet focussing on linear sigma model descriptions.

An outline for the rest of the review is the following. In the next section we will review supergravity aspects of heterotic compactifications. In section \ref{s:worldsheet} we will review some of the lore concerning $(0,2)$ worldsheet SCFTs, their symmetries, and the role of worldsheet instantons. In section \ref{sect:glsm} we will review the $(0,2)$ linear sigma model, its parameter space, quasi-topological twists, several prescriptions for computing correlators and the singular locus of the quasi-topologically twisted theories. We will also outline a proposal for mirror symmetry in a certain class of $(0,2)$ models. In section \ref{s:lg} we will review work aimed at understanding $(0,2)$ Landau-Ginzburg theories. Finally, in section \ref{s:outlook} we will give a brief outlook on open questions in the field.
\vskip 0.5cm
{\bf Acknowledgements}:  It is a pleasure to thank I. Melnikov for helpful discussions and M. Wolf for comments on the manuscript. This work is supported by the EPSRC Postdoctoral Fellowship EP/G051054/1.

\section{Spacetime Aspects of $E_8\times E_8$ Heterotic String Compactifications}
In this section we set the stage for a worldsheet analysis by briefly outlining the supergravity ingredients necessary to define a heterotic string compactification. We will describe some techniques in the simplest examples for computing the spectrum, and Yukawa couplings. Finally, we will comment on some constructions in the literature that give rise to the standard model.

\subsection{Heterotic Effective Field Theory}
At large radius, where all length scales of the target space are large compared to the string length, the dynamics of string theory reduces to that of field theory. The effective field theory describes the interaction of the massless string spectrum, including the metric $G_{MN}$, the $E_8\times E_8$ gauge field $A_M$, its field strength $F_2$ and the heterotic field $B_2$. The action is given by (we largely follow the notation of \cite{Becker:2009df}):
\bea
S &=& \frac{1}{2\kappa^2} \int d^{10}x (-G)^{1/2} e^{-2\Phi}\left[R(\Omega)  + 4|\del_M \Phi|^2 - \half |H_3|^2 - \right.\cr
&&\qquad \qquad\qquad- \left. \frac{\alpha'}{4} \left(\tr |F_2|^2 - \tr |R_2|^2 \right)\right] + {\rm fermions},\label{het_action}
\eea
where $R_2(\Omega_+)$ is the Riemann two-form computed with respect to the spin connection ${\Omega^{PQ}}_M$  twisted by the $H_3$-flux
\bea
{\Omega_{\pm~}^{PQ}}_M =  {\Omega^{PQ}}_M \pm \half {H^{PQ}}_M + \O(\alpha').
\eea
The Einstein-Hilbert term in \C{het_action} is constructed using the spin connection $\Omega$. With this choice of fields and connection, the effective action \C{het_action} is exact to $\O(\alpha')$. It is known how to construct the $\O(\alpha'^2)$ corrections to the heterotic effective action\cite{Bergshoeff:1989de,Bergshoeff:1988nn,Chemissany:2007he}, though such corrections will not be relevant for our purposes.
The last two terms in \C{het_action} are defined as
\bea
\tr |R_2|^2 =\half R_{MNPQ}(\Omega_+) R^{MNPQ}(\Omega_+), \quad \tr |F_2|^2 &=& \half F_{MN} F^{MN},
\eea
and the NSNS field strength is defined as
\bea
H_3 &=& dB_2 +\frac{\alpha'}{4} \left[\CS(\Omega_+) - \CS(A)\right], \label{eqn:H3}
\eea
where CS denotes the Chern-Simons form for the relevant connection. The NSNS field strength obeys a Bianchi identity
\bea
dH_3 = \frac{\alpha'}{4}\left[\tr (R_2(\Omega_+) \w R_2(\Omega_+)) - \tr (F_2 \w F_2)\right].\label{eqn:bianchi}
\eea
Here $\tr (R_2(\Omega_+) \w R_2(\Omega_+))$ is evaluated in the vector representation of $\SO(9,1)$; for the $\Spin(32)/\Z_2$ string, $\tr F_2 \w F_2$ is evaluated in the vector representation; for the $E_8\times E_8$ case (where there is no vector representation) $\tr F_2 \w F_2$ is defined as one thirtieth of the trace in the adjoint representation. The inclusion of the string correction, $\tr (R_2(\Omega_+) \w R_2(\Omega_+))$ is required to cancel anomalies in the underlying string theory, and is required in order to construct solutions with $F_2 \ne 0$.

The conditions of supersymmetry follow from the variation of the fermions in the ten-dimensional effective action. There are three fermions, a gravitino $\Psi_M$, a dilatino $\lambda$ and a gaugino $\chi$. Their variations are given by
\be
\label{sugra_var}
\begin{split}
& \delta \Psi_M  =\Big(\partial_M +\frac{1}{4} {\Omega^{AB}_-}_M
\Gamma_{AB}\Big)\epsilon=0,\cr & \delta \lambda =-\frac{1}{2\sqrt{2}}\Big(  /\!\!\!
\partial \Phi  -{1\over 2}/\!\!\! \!{ H} \Big) \epsilon=0,
\cr  &  \delta \chi  = -{1\over 2} \displaystyle{\not} {F} \epsilon=0
,\cr
\end{split}
\ee
where we have defined the following contractions of $H_3$ and $F_2$:
\be
\Hslash_M = \half H_{MNP}\Gamma^{NP}, \quad \Hslash = \frac{1}{3!} H_{MNP}\Gamma^{MNP},  \quad  \displaystyle{\not} {F}  = \frac{1}{2} F_{MN}\Gamma^{MN}
\ee
As emphasised by \cite{Becker:2009df}, (see also \cite{Hull:1985dx,Kimura:2006af,Sen:1986mg}), the choice of connection in computing the Riemann curvature two-form is important in understanding compactifications with $H_3$-flux. It turns out to be most convenient to choose $\Omega_+$, as this implies the equations of motion and supersymmetry variations remain simple at $\O(\alpha')$ in the $\alpha'$ expansion. With the choice of fields in \cite{Bergshoeff:1989de}, all of the $\O(\alpha')$ corrections to the supersymmetry variations in \C{sugra_var} are contained in the $\alpha'$ modification to $H_3$ in \C{eqn:H3}. One could choose a different choice of connection at the expense of complicating the supersymmetry variations. Finally, for heterotic solutions with type IIB and F-theory duals one naturally generates heterotic solutions with $\Omega_+$ as the preferred connection.

Now that we have established some notation, we wish to explore solutions to this effective action. We are only interested in solutions that preserve $\N=1$ supersymmetry and $d=4$ Poincare invariance.
As such, the ten-dimensional space must take the schematic form
\be
\RR^{3,1} \times M.
\ee
%The metric is taken to be of the form
%\be
%ds^2 = e^{2\phi} \eta_{\mu\nu} dx^\mu dx^\nu + g_{mn} dy^m dy^n,
%\ee
%where $\mu,\nu=0,\ldots,3$ parameterise the Minkowski spacetime, $e^{-2\phi}$ is a warp factor and $g_{mn}$ is the metric on the internal six-dimensional space.
Requiring the $\N=1$ supersymmetry variations in \C{sugra_var} be satisfied vastly simplifies our task of finding solutions. By solving the supersymmetry variations and Bianchi identity we are automatically generating solutions to the equations of motion, though whether the background is a string solution requires tools beyond supergravity. At least in the torsion free case, the supergravity data should be enough to define a sigma model, which generates a $(0,2)$ theory that is at least perturbatively conformal.\footnote{The conditions that supersymmetry be preserved and the Bianchi identity satisfied are sufficient to guarantee the existence of a $(0,2)$ sigma model that is perturbatively conformally invariant \cite{Hull:1986kz,Sen:1986mg}. However, unless one is using a linear sigma model to generate the SCFT, non-perturbative effects may destabilise the vacuum.} If bundle splits non-trivial over every rational curve in $M$ then the theory is argued in \cite{Distler:1987ee} to be conformal non-perturbatively. Alternatively, there is a general belief in the literature, for example \cite{Silverstein:1995re,Beasley:2003fx,Basu:2003bq}, that vacua admitting a linear model description are conformally non-perturbatively, though a comprehensive proof (at least of the former assertion) is still lacking in the literature.

We finish this subsection with a very speculative question: what is the most general $\N=1$ compactification of the heterotic string? This is a notoriously hard question to answer, but whatever the answer is, supergravity solutions are almost certainly a very small corner in the landscape of such vacua. In fact, they should be thought of as a point in a bigger space of string compactifications. This is clear already in $(2,2)$ string compactifications. More generally, the constructions in \cite{McOrist:2010jw} suggest the most general heterotic solutions are most likely non-geometric and have no large volume limit. Thus, in order to understand the general features of heterotic string compactifications, it is necessary to have a worldsheet understanding. There are many known examples of string backgrounds without supergravity limits. Perhaps the most familiar class of examples are compactifications described by Landau-Ginzburg theories and asymmetric orbifold compactifications. These have no supergravity or geometric interpretation, and are defined purely in terms of their worldsheet theory. A second class of backgrounds, known as non-geometries, are those that are patched together using T-duality or other quantum symmetries of string theory (see \cite{Wecht:2007wu} for a review and references therein). Even in the heterotic string (where there are no RR fluxes), a correct worldsheet description of these backgrounds is not yet well-understood.\footnote{ One proposal is that of a doubled-torus formalism of Hull (for example \cite{Hull:2009sg}), while the second is given by wrapping the tensor theory describing an M5-brane on a K3 surface \cite{McOrist:2010jw,Sethi:2007bw}.} Finally, a third class of compactifications without large volume limits, are the torsional heterotic solutions (see for example\cite{Dasgupta:1999ss,Becker:2003yv,Becker:2003gq,Becker:2006et,Becker:2003sh,Becker:2009zx,Becker:2002sx,Becker:2009df,Becker:2005nb,Becker:2008rc,Goldstein:2002pg,Andreas:2010qh,Carlevaro:2009jx,Carlevaro:2008qf,Fu:2008ga,Becker:2006xp,Fu:2006vj}). It is thus clear that to have a conclusive understanding of string compactifications, one needs to have an understanding of the string worldsheet. We will restrict to compactifications that have a supergravity limit. Even with this vast simplification, the worldsheet structure of such solutions is rich and intricate.

\subsection{Heterotic Vacua with K\"ahler Metrics}
\label{ss:kahler}
We will now simplify the discussion to $E_8\times E_8$ heterotic vacua with large radius limits. What are the geometric properties of such solutions? By analysing the supersymmetry conditions \C{sugra_var}, and anomaly cancellation, it was shown by \cite{Strominger:1986uh} that such manifolds are complex, with vanishing first Chern class and obey the equations
\bea
2i\del\delbar J &=&  \frac{\alpha'}{4}\left[\tr (R_2(\Omega_+) \w R_2(\Omega_+)) - \tr (F_2 \w F_2)\right],\cr
F^{(0,2)} &=& F^{(2,0)} = 0, \quad F_{MN}J^{MN}=0,\cr
d(e^{-2\Phi} J \w J) &=& 0.\label{eqn:torsion}
\eea
Here $J$ is the hermitian form for the target space. The first equation implies the space is non-K\"ahler if $H_3\ne 0$. Let us analyse their import order-by-order in an expansion of $1/r$ under rescalings of the coordinates, where $r$ is the characteristic radius of $M$. As the gauge field $A_M$ appears in the covariant derivative $D_M = \del_M + i A_M$, we take $A_M = \O(1/r)$. To $\O(1/r)$ the gravitino variation implies that $M$ is a Calabi-Yau manifold with $\SU(3)$ holonomy, while the dilatino variation implies a constant dilaton.  The gaugino variation is a $\O(1/r^2)$ constraint and implies the bundle $\cF$ must satisfy the constraints
\bea
F^{(0,2)}=F^{(2,0)}=0,\label{Eholo} \\
F_{MN}J^{MN}=0.\label{hym}
\eea
The first equation implies the bundle is holomorphic (i.e. all the transition functions are holomorphic functions with respect to the complex structure on $M$), while the last equation is known as the hermitian Yang-Mills equation. The first two equations are relatively straightforward to satisfy, while the last equation is notoriously hard to solve. Fortunately, on K\"ahler manifolds there is a way of turning this into more tractable problem using the Donaldson-Uhlenbeck-Yau theorem. This theorem states that for a holomorphic vector bundle $\cF$ on a K\"ahler manifold with a given complex structure, there exists a unique connection satisfying \C{hym} provided the bundle $\cF$ is ``stable.'' We will not go into the details of this theorem, or what stability specifically means, apart from the fact it is a rather mild quasi-topological constraint. 

%(Strictly speaking a bundle may be poly-stable: it is a direct sum of stable bundles satisfying a certain constraint).

 %and we have allowed for the possible presence of NS5-branes wrapping holomorphic curves represented by the class $W_5$.\footnote{ Classes that have a holomorphic curve representation are known as effective.}
%%Stability of a sheaf $\F$ is defined via its slope
%\be
%\mu (\F) = \frac{1}{{\rm rk} (\F)} \int_X c_1(\F) \w J \w J,
%\ee
%where $J$ is the K\"ahler form on $M$, and ${\rm rk}(\F)$ and $c_1(\F)$ are the rank and first Chern class of $\F$. A bundle is stable if every sub-sheaf $\F\subset\cF$ satisfies $\mu(\F) < \mu(\cF)$. (A bundle is semi-stable if the inequality is not strict). A bundle is poly-stable if it is decomposable into a direct sum of stable bundles all with the same slope.

\subsubsection{The Standard Embedding}
The simplest solution with a K\"ahler metric at large volume is known as the standard embedding. In this section we will review some of the pertinent features --- more details may be found in say \cite{Polchinski:1998rr,Green:1987mn}. The solution proceeds by assuming that $dB_2=0$. The Bianchi identity then reduces to
\be
0 = \tr (R_2(\Omega_+) \w R_2(\Omega_+)) - \tr (F_2 \w F_2).
\ee
This equation is essentially impossible to solve unless there is a special relation between $F_2$ and $R_2(\Omega_+)$. Such a relation is given by identifying the spin connection with the gauge connection. In that case $R_2(\Omega_+) = F_2$, and the Bianchi identity is satisfied identically. Such a gauge choice satisfies the hermitian Yang-Mills equations, is supersymmetric, and is a well-defined starting point for defining a sigma model description. If $M$ has $\SU(3)$ holonomy (and not a subgroup), then the bundle will have $\SU(3)$-structure. The $\SU(3)$ is embedded in one of the $E_8$ gauge groups, with the other $E_8$ gauge group regarded as the hidden sector. The unbroken gauge symmetry is given by the commutant of $\SU(3)$ with the $E_8$, which turns out to be $E_6$. Thus, without much work we have constructed a perturbative string compactification that has an unbroken $E_6$ gauge group. This is to be contrasted with the recent work in F-theory model building, in which $E_6$ gauge groups can only be generated non-perturbatively (see for example \cite{Beasley:2008dc}).

The string compactification is well described by a $d=4$, $\N=1$ effective field theory when the length scales are well above the compactification scale $r$, but well-below the string length $l_s$. The massless field content and  interactions of this field theory may be constructed by Kaluza-Klein reducing the $d=10$, $\N=1$ supergravity theory on the internal Calabi-Yau manifold. As there is an unbroken $\N=1$ supersymmetry, the theory is most conveniently represented in terms of $\N=1$ superspace, and we will describe, somewhat schematically, its field content and construction. We will mostly follow the notation and discussion in \cite{Polchinski:1998rr,Green:1987mn}.

The field content consists of Kaluza-Klein reducing $G_{MN}$, $B_{MN}$, $\Phi$ and $A_M$ on $M$. There are $h^{2,1}$ chiral multiplets for the complex structure moduli, whose bosonic fields are $G_{ij},G_{\bar i \bar j}$ (here $i,j$ are holomorphic indices on $M$); there are $h^{1,1}$ chiral multiplets corresponding to the complexified K\"ahler moduli given by $G_{i\bar j} + B_{i \bar j}$.

The gauge group splits as $E_8 \times E_8 \rightarrow \SU(3)\times E_6 \times E_8$, giving rise to an unbroken $E_6$ gauge group in spacetime (together with an $E_8$ hidden sector). Kaluza-Klein reducing the $d=10$, $E_8\times E_8$ gauge multiplet gives rise to some matter that is charged under the $E_6$, as well as some matter that is uncharged. To see this note that the adjoint gauge field $A_{M}$ decomposes as
\begin{equation}({\bf 248},{\bf 248}) \rightarrow ({\bf 1},{\bf 78},{\bf 1})+({\bf 1},{\bf 1},{\bf 248})+({\bf 3},{\bf 27},{\bf 1})+({\bf \bar 3}, \tsevbar, {\bf 1}) + ({\bf \bar 8}, {\bf 1}, {\bf 1}).
\end{equation}
The first two components, denoted by the field $A_{\mu,a}$ correspond to the $d=4$, $E_6\times E_8$ massless gauge field with $a$ labelling the adjoint of $E_6\times E_8$. Together with their fermionic partners, these fields form a $d=4$, $N=1$ gauge multiplet. As we have identified the $\SU(3)$ spin connection with the $\SU(3)$ gauge connection, the charged matter content comes from the component of the gauge field transforming as the $({\bf 3},{\bf 27},{\bf 1})+({\bf \bar 3}, \tsevbar, \bar {\bf 1})$, and are given by the fields $A_{i,jx}$ and $A_{\bar i,\bar j\bar x}$, where $x$ is the ${\bf 27}$ of $E_6$. By contracting with the holomorphic 3-form, the former corresponds to a $(2,1)$-form giving rise to $h^{2,1}$ chiral multiplets transforming in the ${\bf 27}$ of $E_6$. The latter gives rise to $h^{1,1}$ chiral multiplets transforming in the $\tsevbar$ of $E_6$. Finally, there are a number of gauge singlets $A_{i,j\bar k}$ transforming in the adjoint of $\SU(3)$. These are 1-forms valued in the endomorphism group of the tangent bundle, and correspond to elements of the cohomology group $H^1(M,{\rm End} (TM))$. They parameterise classically flat directions in which $\cF$ is deformed from the tangent bundle to a more general $\SU(3)$ vector bundle. We will discuss this further in the next subsection.

In the above, we implicity used supersymmetry to deduce the gaugino decomposed in the same way as the gauge field. There are thus $h^{2,1}$ chiral multiplets in the ${\bf 27}$ of $E_6$, and $h^{1,1}$ chiral multiplets in the $\tsevbar$ of $E_6$. The spectrum is therefore chiral with the net number of generations given by
\be
N_{\rm gen} = |h^{2,1} - h^{1,1}| = \frac{|\chi|}{2},
\ee
where $\chi$ is the Euler characteristic of $M$. This can also be understood using an index theorem for the Dirac operator. $N_{\rm gen} = 100$ for the quintic in $\P^4$, and $N_{\rm gen} = 4$ for the $\Z_5\times \Z_5$ orbifold of the quintic. Thus even in the simplest example it is not hard to get down to a reasonable number of generations. However, it is quite hard to find Calabi-Yau manifolds with $|\chi(M)|=6$, with only a handful of known examples. For compactifications involving gauge bundles that differ from the standard embedding, the number of generations is no longer tied to the Euler characteristic, instead it depends on topological quantities associated with $\cF$ and $M$. This allows for more phenomenological flexibility.

Having described the field content of the $d=4$ effective field theory, we will now describe the Lagrangian that governs their interactions. To specify the Lagrangian we need to construct the K\"ahler potential and superpotential for all of the fields. As the Calabi-Yau geometry solves the string equations of motion for any value of the moduli fields, the moduli do not have any superpotential terms.
%This is believed to hold non-perturbatively in $\alpha'$ for the complex structure and K\"ahler moduli; worldsheet instantons are known to generate a potential for the gauge singlets $H^1({\rm End}(TM))$, and as such, will be discussed in the next section.
The   $\tsev$ and $\tsevbar$ matter fields are not moduli and have a non-trivial superpotential. They are both massless, and the vacuum is stable (no tadpole) so the lowest order term is a cubic interaction.
%This Yukawa interaction comes from reducing the $d=10$ Yang-Mills Yukawa interaction and can be represented in $d=4$ $\N=1$ superspace as a superpotential.
Although we will not give any details of the proof originally derived in \cite{Dixon:1990pc}, it turns out that the low-energy dynamics of the light fields is tightly constrained. The two derivative dynamics of the moduli and the matter fields are intricately related to each other, and their interactions are completely determined in terms of two holomorphic functions. These relations are known as special geometry and are a consequence of certain properties of  the worldsheet theory. This relation is known to hold for type II theories and for heterotic compactifications with the standard embedding. For a summary of the local geometry of the moduli space see  \cite{Candelas:1990pi}. It is an interesting open question as to whether it holds for more general embeddings in the heterotic string. Discussing these relations is thus beyond the scope of this review.

While the standard embedding has many nice features: an $E_6$ GUT group, special geometry and a well-understood worldsheet description, there are also many phenomenological problems. For example, in a GUT scenario, where say $E_6$ is broken to $\SU(5)$ and then to the standard model via Wilson lines, the $\tsev$ of $E_6$ gives rise to particles transforming as the ${\bf 5} +\overline{{\bf 5}}$ of $\SU(5)$. These particles are yet to be observed at current energy scales and mediate rapid proton decay---at least without a mechanism to give these particles a large mass, this is a phenomenological problem. It is also hard to give neutrinos a mass using the usual see-saw mechanism in $E_6$ compactifications. The required Yukawa coupling which are forbidden in string perturbation theory (see for example \cite{Dine:1985vv}). It is thus desirable to understand more general heterotic compactifications, and the easiest way to achieve this is to consider more general gauge bundles. For example $\SU(4)$ and $\SU(5)$ bundles give rise to unbroken $\SO(10)$ and $\SU(5)$ gauge groups, which alleviate some of these problems. However, although one can construct such compactifications in supergravity, it is not known how to construct their worldsheet descriptions.

\subsubsection{General Gauge Bundles}
It is phenomenologically desirable to consider compactifications beyond the standard embedding. Recall that in order to preserve supersymmetry, the bundle $\cF$ must be holomorphic and satisfy the hermitian Yang-Mills equations. Equivalently, if one has a stable bundle, then the Donaldson-Uhlenbeck-Yau theorem guarantees the existence of a unique connection solving these equations. A lot of sophisticated mathematical technology has been developed over the years to construct stable bundles and hence supergravity solutions. One is based on the Friedman-Morgan-Witten constructions \cite{Friedman:1997yq,Friedman:1997ih,Friedman:1998si,Braun:2005zv,Braun:2005ux} for elliptically fibered Calabi-Yau manifolds. Another is based on complete intersection Calabi-Yau manifolds with monad bundles\cite{Distler:1987ee,Blumenhagen:2006wj,Blumenhagen:1997vt,Kachru:1995em}. Some recent examples in the context of heterotic model building include \cite{Anderson:2009mh,Andreas:2007ei,Anderson:2007nc,Buchmuller:2005jr,Braun:2005nv,Bouchard:2005ag,Braun:2005ux,Blumenhagen:2010ed,Anderson:2010mh} and a review of the techniques used to computing the low-energy physics of such compactifications is given in \cite{Anderson:2008ex}.   Although such constructions can give rise to higher rank bundles, and are phenomenologically desirable, they are typically restricted to the regime of supergravity and it is not known how to describe their compactifications using worldsheet techniques. Such techniques would be very useful.  For example, one may be able to construct the necessary Yukawa couplings for generating neutrino masses using non-perturbative corrections. Thus, we will focus on the latter class of bundles where there has been recent work in constructing well-defined worldsheet descriptions, and a certain class of unnormalised Yukawa couplings can be computed exactly in $\alpha'$. It will be interesting in future work to extend these results to the more general bundles, in particular, those not attainable as deformations from the standard embedding.

\subsubsection{Beyond Supergravity and Onwards to a Worldsheet Description?}
At large radius, the moduli corresponding to deforming the tangent bundle come from the gauge singlets corresponding to elements of the cohomology group $H^1(M,{\rm End} (TM))$. The Bianchi identity \C{eqn:bianchi} implies there is non-vanishing torsion $H_3$, and the worldsheet theory has a reduced amount of supersymmetry: $(2,2)$ broken down to $(0,2)$. Early on, starting with \cite{Dine:1986zy}, it was thought that these bundle moduli were lifted by non-perturbative $\alpha'$ effects. Worldsheet instantons wrapping holomorphic curves in $(0,2)$ theories have a reduced number of zero modes, and generically generate a potential for these moduli, thereby destabilising the vacuum (we expand on this phenomenon in section \ref{ss:dest} below). However, for certain compactifications, including those connected to the standard embedding, such instantons are known to be either entirely absent \cite{Distler:1986wm} or to cancel amongst themselves\cite{Beasley:2003fx,Silverstein:1995re,Basu:2003bq}. Consequently, the deformations parametrised by
$H^1(M,{\rm End} (TM))$ are unobstructed and are genuine moduli.

Nonetheless, it is impossible to deduce that these deformations are true moduli purely in supergravity. It is worldsheet instantons that destabilise the vacuum, and to construct vacua that are free of these instanton effects, it is necessary to have a worldsheet description of such vacua. Furthermore, $\alpha'$ quantum corrections are known in other context to generate qualitatively new phenomenon not seen at large volume. One example is mirror symmetry, in which strong coupling effects are exchanged with weak coupling effects. Another example is quantum cohomology, in which the cohomology of a Calabi-Yau $M$ is generalised to include quantum corrections. What is the generalisation of quantum cohomology to spaces with vector bundles? Do $\alpha'$ corrections qualitatively modify Yukawa couplings? Is there a notion of special geometry for heterotic compactifications with gauge bundles that are not the standard embedding? Recently there has been a body of work aimed at answering those question, with which we will summarise in the remainder of this review.

\section{The Worldsheet: (0,2) SCFTs and the GLSM}
\label{s:worldsheet}
A perturbative string compactification is defined via an anomaly free SCFT with the correct central charge. If the background has a geometric interpretation, then the SCFT may be generated by following the RG flow of a non-linear sigma model. However, the non-linearity makes it hard for one to compute anything concretely. A more tractable approach is to appeal to universality: perhaps there is an easier model to study in the same universality class. For Calabi-Yau's defined as complete intersections in toric varieties with monad bundles there is a very convenient class of field theories that do this job: gauged linear sigma models (GLSMs). These are abelian gauge theories with linear kinetic terms coupled by a chiral and twisted-chiral superpotentials. They provide a useful probe of the SCFT moduli space, and can compute certain terms in the $d=4$ low-energy effective field theory. Their utility is well-established in the world of $(2,2)$ SCFTs.  Here we outline progress in using GLSMs to study $(0,2)$ SCFTs. As first step we will review some basics of the non-linear sigma model, before turning to the linear sigma model in the next section.

\subsection{Non-linear Sigma Models: Target Space versus Worldsheet}
The starting point for describing a perturbative heterotic string compactification is a two-dimensional non-linear sigma model, a theory of maps from a Riemann surface to a ten-dimensional target space. The field content consists of the maps \mbox{$X^M:\Sigma\rightarrow M$}
 for $M=0,\ldots,9$; a number of right-handed world sheet fermions $\psi^M$, which couple to the pullback to $\Sigma$ of the tangent bundle of $M$; and a number of left-handed worldsheet fermions $\gamma^A$, which couple to the pullback of the vector bundle $\cF$. For the $E_8\times E_8$ heterotic string, the bundle splits into two $E_8$ bundles $\cF_1,\cF_2$, of which we will always regard $\cF_2$ as trivial---it is regarded as the unbroken $E_8$ gauge group of a hidden sector. In the following we will often write $\cF$ for $\cF_1$, dropping the subscript. In two-dimensions, the supersymmetries are Majorana-Weyl and are chiral: we can have independent numbers of left-handed and right-handed supersymmetries. These are labelled $(N_L,N_R)$, so that $(0,2)$ implies there are two right-moving supersymmetries and no left-moving supersymmetries. We must always have at least $(0,1)$ supersymmetry to have a consistent string background, while if there is $N=1$ spacetime supersymmetry then there is at least $(0,2)$ worldsheet supersymmetry. When $\cF = TM$ with the gauge and spin connections identified, the right-moving and ten of the left-moving fermions transform in an identical fashion. Consequently, the theory has $(2,2)$-supersymmetry.

The sigma model is action is given by
\bea
S &=& \frac{1}{4\pi\alpha'}\int d^2z \left\{[G_{MN}(X) + B_{MN}(X)]\del_z X^M \del_{\bz} X^N + G_{MN}(X) \psi^M \nabla_z \psi^N + \right.\cr
&& \left.\quad+ \gamma^A \nabla_{\bz} \gamma^A + \half F_{\rho\sigma}^{AB} \gamma^A \gamma^B  \psi^\rho  \psi^\sigma\right\},\label{eqn:sigma_model}
\eea
where $A,B=1,\ldots,32$ label the bundle indices and as usual, we have fixed to conformal gauge---for more details see \cite{Polchinski:1998rr}. The covariant derivatives are
\bea
\nabla_z \psi^M &=& \del_z \psi^M + \left[\Gamma^M_{~NP}(X) - \half H^M_{~NP}(X)\right] \del_z X^N \psi^Q,\cr
\nabla_\bz \gamma^A &=& \del_{\bz}\gamma^A - i A_{M}^{AB}(X) \del_{\bz} X^M \gamma^B,
\eea
with $A_M$ the connection on $\cF$. The spacetime fields such as the metric, B-field and connection act as worldsheet couplings, and are in principle complicated functions of the $X^M$ fields. This makes the theory highly non-linear and generally hard to solve (that is, one cannot easy deduce the spectrum, or correlation functions). Furthermore, for a generic choice of target space fields the theory is not conformal---a necessary requirement to define a critical string theory. These two basic issues tend to hamper a direct study of non-linear sigma models. However, one can make progress by considering a large volume limit, in which the coupling of the non-linear sigma model is ``small'' and a semi-classical analysis is valid. One can then extract the spectrum, compute scattering amplitudes and compute the beta function perturbatively in $\alpha'$, and in some cases extract all order results in $\alpha'$. For example, there may exist a truncation of the non-linear sigma model to its `topological' sector, in which the theory becomes completely independent of the worldsheet metric (for example \cite{Witten:1988xj}).

The sigma model has diffeomorphism anomalies (i.e. the sigma model is not modular invariant) unless $M$ and $\cF$ satisfy certain topological constraints (in integer cohomology):
\bea
c_1(M) &=& 0, \quad c_1(\cF) = 0 ~~{\rm mod}~2, \label{anomaly2} \\
\ch_2(\cF) &=& \ch_2(M),\label{anomaly3}
\eea
The first condition \C{anomaly2} implies $M$ admits spinors, as this amounts to the second Stiefel-Whitney class of $M$ vanishing. This is physically very reasonable: we want to be able to define spinors on our spacetime, and furthermore, one needs both the vector and spinor representations of $\SO(16)$ in order to recover the full $E_8$ representation. The second equation \C{anomaly3} can be recast in the familiar form of the Bianchi identity for the three-form field strength viz. \C{eqn:bianchi}.

If the anomalies are cancelled, then one is free to couple the theory to worldsheet gravity and to gauge fix the resulting two-dimensional supergravity theory. However, it must also be checked that the conformal mode of the string vanishes, so that the longitudinal modes of the string decouple in scattering amplitudes. In the large volume limit, this may be checked perturbatively in $\alpha'$. At one-loop, the vanishing of the beta function implies the space $M$ is Ricci flat and that the bundle satisfies the hermitian Yang-Mills equations (c.f. \C{hym}). Hence, the combination of Yau's theorem and the Donaldson-Uhlenbeck-Yau theorem imply the existence of a solution to the beta function at 1-loop. Higher order loops will not affect the topological conditions, while the geometrical conditions of Ricci-flatness and the hermitian Yang-Mills equations \C{hym} will be corrected order by order in $\alpha'$ perturbation theory. The quantum corrections are special though; the sigma model re-adjusts itself order-by-order in $\alpha'$ perturbation theory to ensure the existence of a supersymmetric solution\cite{Witten:1985bz,Witten:1986kg}.\footnote{This is only true if $X:\Sigma\rightarrow M$ is a topologically trivial map; worldsheet instantons invalidate this argument. These effects will be discussed later in this review.} The lack of conformal invariance may be viewed as renormalisation group flow, in which degrees of freedom become massive and decouple in the infrared. This leaves us with a finite number of degrees of freedom that characterise the massless spectrum of the conformal field theory. These degrees of freedom tend to be topological in nature and include the complex structure of $M$; the holomorphic structure of the vector bundle $\cF$; and the cohomology class of the complexified K\"ahler class. Nonetheless it is quite hard to analyse these conformal field theories directly, and as such one must resort to other methods to which we now turn.

\subsection{The Symmetries and Spectrum of $(0,2)$ Theories}
\label{symm}
We now briefly analyse the symmetries and spectrum of $(0,2)$ non-linear sigma models. This will serve to further introduce some notation, and will naturally lead into a study of $(0,2)$ gauged linear sigma models.

The part of the theory involving $X,\psi$ associated with the flat spacetime, and $\gamma$'s that do not couple to the gauge bundle (for example the hidden $E_8$) are free fields and analysing their behaviour is straightforward. The rest of the theory is a non-linear sigma model, which is interacting and hard to get a handle on. One approach is to follow the work of \cite{Rohm:1985jv} by using the Born-Oppenheimer approximation in which we truncate theory to quantum mechanics. This was applied in the context of $(0,2)$-theories in \cite{Distler:1987ee}, (see also \cite{Distler:1995mi}) and provides a useful route to get a handle on the physics the sigma model. An alternative approach is to appeal to universality and construct an abelian gauge theory (that is, a gauged linear sigma model) that under RG flows to the same infrared fixed point. In this subsection we will content ourselves with an analysis of the non-linear sigma model, before turning in the next section to GLSMs.

As mentioned above, the requirement of $\N=1$ spacetime supersymmetry implies at least $(0,2)$ worldsheet supersymmetry. The left-moving sector of the worldsheet has a Virasoro symmetry, and we will require an additional $\U(1)_L$ symmetry
\be
J(z) J(0) \sim \frac{r}{z^2}.
\ee
where $r$ is a parameter of the $\U(1)_L$ algebra and will eventually correspond to the rank of the gauge bundle. There are a number of reasons for requiring this $\U(1)_L$ symmetry. Firstly, it provides a natural candidate for a chiral GSO projection on the left-moving fields. It acts as
 \be
 g = e^{-i\pi J_0} (-1)^{F_L},
 \ee
 where $F_L$ is the left-moving fermion number.  Secondly, the $\U(1)_L$  plays an important role in realising the spacetime gauge symmetry. Thirdly, it appears naturally when one considers gauge bundles obtained as deformations from the standard embedding and will occur in the class of compactifications we consider in section \ref{sect:glsm} onwards.

The $X^M$ bosons associated with $\RR^{3,1}\times M$ and $r$ Weyl fermions associated with the gauge bundle give a left-moving central charge of $c_L = 10+r$. To cancel the anomaly we need $c_L = 26$. As $r\ne0$, one can do this by adding some free Majorana-Weyl fermions $\gamma^I$ for $I=1,\ldots,16-2r$, and $16$ free fermions $\wt \gamma^A$, $A=1,\ldots,16$, associated with the hidden $E_8$. The $\U(1)_L$ acts on the fermions analogously to the $\U(1)_R$ on the $\psi^M$:
\be
\gamma^{i} \rightarrow e^{i\theta_L} \gamma^{i}, \quad \gamma^{\bar i} \rightarrow e^{-i\theta_L} \gamma^{\bar i}.
\ee
The theory then linearly realises a $\U(1)\times \SO(16-2r)\times \SO(16)$ spacetime gauge group. The $\U(1)_L$ also provides a left-moving spectral flow generator: when acted on NS states created by the $\gamma^I$ it takes them to R ground states and vice-versa. Thus, even though only $\U(1)\times \SO(16-2r)$ of the spacetime gauge group is linearly realised, there are additional states (the vertex operators for which may be explicitly constructed by bosonisation) implying the spacetime gauge group is actually larger than $\U(1)\times \SO(16-2r)$.\footnote{The same phenomenon occurs for the hidden sector $E_8$, where only the $\SO(16)$ subgroup is linearly realised. A naive computation of the spectrum in flat space gives states in the $120+128$ representation of $\SO(16)$. Consistency of the spacetime theory however implies the gauge bosons must be in the adjoint representation of the gauge group. There is precisely one group, $E_8$, under which the adjoint decomposes into $120+128$. The remaining states in the adjoint representation may be constructed using spin fields and bosonisation. Thus, the full spacetime gauge group is $E_8$. } In fact, for $r=3,4,5$ the full spacetime gauge group is  $E_6\times E_8$, $\SO(10)\times E_8$ and $\SU(5)\times E_8$ respectively. We have listed the linear representations and how they assemble themselves into representations of the full gauge group for $r=4,5$ in table~\ref{table:reps}.

\begin{table}[t]
\begin{center}
\begin{tabular}{|c|c|c|}
\hline
${\rm Rep.~of~}E_6   	$&${\rm Rep.~of~}\SO(10)\times\U(1)   	$&${\rm Cohomology~Group}$\\ \hline
${\bf 78} $&$ 	{\bf 45}_0\oplus{\bf 16}_{-3/2}\oplus{\ol{\bf 16}}_{3/2}\oplus {\bf 1}_0	$&$H^*(M,\O)$\\ \hline
${\bf 27} $&${\bf 16}_{1/2}\oplus{\bf 10}_{-1}\oplus {\bf 1}_0                             			 $&$H^*(M,\cF)$\\ \hline
${\bf 1}  $&${\bf 1}_0                                       			$&$H^*(M,{\rm End}\, \cF)$\\ \hline
\end{tabular}
\vskip1cm
\begin{tabular}{|c|c|c|}
\hline
${\rm Rep.~of~}\SO(10)   	$&${\rm Rep.~of~}\SO(8)\times\U(1)   	$&${\rm Cohomology~Group}$\\ \hline
${\bf 45} $&${\bf 8}_{-2}^{s'}\oplus {\bf 28}_0\oplus{\bf 1}_0\oplus{\bf 8}_2^{s'}			$&$H^*(M,\O)$\\ \hline
${\bf 16} $&${\bf 8}_{-1}^{s}\oplus {\bf 8}_1^v             			$&$H^*(M,\cF)$\\ \hline
${\bf 10} $&${\bf 1}_{-2}\oplus {\bf 8}^{s'}_0\oplus{\bf 1}_2                 			$&$H^*(M,\bigwedge^2\cF)$\\ \hline
${\bf 1}  $&${\bf 1}_0                                          			$&$H^*(M,{\rm End}\, \cF)$\\ \hline
\end{tabular}
\vskip1cm
\begin{tabular}{|c|c|c|}
\hline
${\rm Rep.~of~}\SU(5)   	$&${\rm Rep.~of~}\SO(6)\times\U(1)   	    $&${\rm Cohomology~Group}$\\ \hline
${\bf 24}           $&${\bf \ol{4}}_{-5/2}\oplus {\bf 15}_0\oplus{\bf 1}_0\oplus{\bf 4}_{5/2}			 $&$H^*(M,\O)$\\ \hline
${\bf 10 }          $&${\bf 4}_{-3/2}\oplus {\bf 6}_1             			        $&$H^*(M,\cF)$\\ \hline
${\bf \overline{5}} $&${\bf \overline{4}}_{-1/2}\oplus {\bf 1}_2                 			 $&$H^*(M,\bigwedge^2\cF)$\\ \hline
${\bf 1}  $&${\bf 1}_0                                          			$&$H^*(M,{\rm End}\, \cF)$\\ \hline
\end{tabular}

\end{center}
\caption{The representations of the spacetime gauge group and how they decompose into their linearly realised subgroups. See \cite{Distler:1987ee} for the case with general $r$.}
\label{table:reps}
\end{table}

\subsection{Instantons and Vacuum Destabilisation}
\label{ss:dest}
A string compactification typically requires conformal invariance of the sigma-model. The vanishing of the 1-loop beta function implies certain conditions of the target space, for example $M$ must have a Ricci-flat metric. These conditions are satisfied for the $N=1$ compactifications described in section \ref{ss:kahler}. However, the conformality conditions are modified at higher loop orders, and we do not know what the all-orders condition of conformal invariance is. A more fruitful approach is to view the lack of conformal invariance as renormalisation group flow---the metric and other target-space fields are quantum corrected order-by-order in $\alpha'$. In that case, the question one must ask is: can the metric and field content be consistently modified such that at each order in $\alpha'$, the equations of motion and supersymmetry are satisfied? As discussed in \cite{Witten:1985bz,Witten:1986kg}, the answer, at least in $\alpha'$ perturbation theory, comes from the non-renormalisation the spacetime superpotential. The essence of the argument comes from holomorphy and the presence of a Peccei-Quinn symmetry. Holomorphy implies perturbative corrections must appear in terms of the complexified K\"ahler modulus $t=a + i V$, the complex combination of the volume modulus $V$ and the modulus $a$ coming from the B-field. The role of the Peccei-Quinn type symmetry is to forbid any non-derivative couplings involving $a$, implying $a$ decouples at zero momentum. This is may be seen by studying the vertex operator associated to $a$:
\be
V_a(k) \sim \int_\Sigma B_{m\bar n}(X) \left[(\del_z X^m + ik\cdot\psi \psi^m)\del_\bz X^{\bar n} - (\del_z X^{\bar n} + i k\cdot \psi \psi^{\bar n}) \del_{\bz} X^m \right]e^{i k\cdot X}.
\ee
The indices $(m,n)$ are holomorphic indices along the Calabi-Yau $M$. At zero-momentum it takes the form
\be
V_a(0) \sim \int_\Sigma B_{m\bar n}(X) \del_z X^m \del_\bz X^{\bar n},\label{eqn:b}
\ee
and is interpreted as the pull-back of the two-form $B_2$ via $X$ to the worldsheet $\Sigma$. If $X$ is a topologically trivial map, as is the case in perturbation theory, then this automatically vanishes. Hence $a$ decouples at zero-momentum, and the four-dimensional effective theory has a symmetry $a \rightarrow a + {\rm constant}$, giving $a$ an axion-like behaviour. This implies $a$ cannot appear in the spacetime superpotential, and there are no perturbative corrections in $t=a+i V$. Consequently, if we have solved the supersymmetry conditions at large radius, they will be automatically satisfied at all orders in $\alpha'$ perturbation theory. As hinted at above, even though our starting point was a Calabi-Yau metric with constant dilaton and no $H_3$, these will become non-zero at higher orders in $\alpha'/r^2$ in such a way to preserve supersymmetry.

This argument is violated by worldsheet instantons, in which $X$ is topologically non-trivial\cite{Dine:1987bq}. The instanton action is given by the pullback of the metric to $\Sigma$ via $X$:
\be
S_{\rm inst} \simeq i \int_\Sigma G_{m\bar n} (\del_z X^m \del_{\bz} X^{\bar n} + \del_\bz X^m \del_{z} X^{\bar n} ).\label{eqn:inst}
\ee
Using K\"ahlerity of $G_{m\bar n}$, and the fact the instanton corrections to the spacetime superpotential must be holomorphic in $t$, one can show that stationary points of this action occur when 
\be
\del_\bz X^m = 0.
\ee
In this case the instanton action saturates a minimal bound given by $S_{\rm inst} = |\int_\Sigma V_a(0)|$. Finally, the instanton must only contribute at genus zero (i.e. on the sphere), as the superpotential does not receive any corrections from finite orders in string perturbation theory\cite{Dine:1986vd}.

The corrections to the superpotential which are potentially dangerous to the stability of the vacuum are of the form
\be
W\sim e^{2\pi i T} \times {\rm constant},\label{eqn:inst_2}
\ee
where $T$ is the complex superfield corresponding to the scalar fields $a+iV$. Such contributions are computed via scattering amplitudes of vertex operators that are gauge singlets (for example, elements of $H^1(M,{\rm End}TM)$), and consequently do not couple to the free left-moving fermions $\gamma^I$. Generically, there exist holomorphic curves of genus zero in Calabi-Yau manifolds and it would appear, as originally argued in \cite{Dine:1986zy}, that scattering amplitudes of the form \C{eqn:inst_2} are non-zero and the vacuum is unstable. However, it is possible that instantons of the form \C{eqn:inst_2} vanish. When could this occur? As pointed out in \cite{Distler:1987ee},  one possibility is that there are $\gamma^I$ zero-modes in the instanton background. If this is the case, then the scattering amplitude must vanish, as the gauge singlets do not couple to the $\gamma^I$. Such zero-modes occur in $(2,2)$ compactifications, as the left-moving supersymmetry can simply act on the bosonic zero modes of the instanton. (This is why $(2,2)$-compactifications are stable.) Another possibility is that instantons are present, but when summed up magically give zero.

In the former case, the precise conditions for when one expects $\gamma^I$ zero-modes to occur are fleshed out in \cite{Distler:1986wm,Distler:1987ee} which we parrot here. First, we need to use a theorem of Grothendieck which states that any holomorphic vector bundle on a two-sphere splits as a direct sum of holomorphic line bundles. Thus, for a worldsheet instanton wrapping a holomorphic curve in $M$, the vector bundle pulls back to $\Sigma$ as a direct sum of line bundles. Thus the left-moving fermions $\gamma^I$ are coupled to a direct sum of line bundles. Line bundles on $\PP^1$ are classified by a single integer, the integral of their first Chern-class, and are denoted $\O(n)$, where $n$ is the relevant integer. The tangent space to $\PP^1$ is $\O(2)$ while the spin bundle is $\O(-1)$. The Dirac operator for the $\gamma^I$ fermions coupled to a vector bundle $\cF$ is given by $\delbar_{\cF\otimes\O(-1)}$. This implies zero-modes of $\gamma^i$ correspond to elements of $H^0(\cF\otimes \O(-1))$, while zero-modes of $\gamma^{\bar i}$ correspond to elements of $H^1(\cF\otimes \O(-1))$. By Grothendieck's theorem,
$$  \cF = \oplus_i \O(n_i),$$
where $\sum n_i = c_1(\cF)=0$ (by anomaly cancellation considerations). Zero-modes are counted by the dimensions of the relevant cohomology groups, which are given by
\bea
h^0\left(\O(n)\right) &=& \left\{
\begin{array}{lr}
k+1 & k\ge 0\\
0 & k\le -1
\end{array}\right.,\cr
h^1(\O(n)) &=& \left\{
\begin{array}{lr}
-k-1 & k\le-2\\
0 & k\ge -1
\end{array}\right.,
\eea
Note that $h^0(\O(-1))$ and $h^1(\O(-1))$ are both non-zero, so there is a chance of having fermionic zero-modes. In fact, if any of the $n_i$ are non-zero then there are necessarily $\gamma$ zero-modes. The argument just given only applies for single-instantons, but is easily extended to multi-instantons. If there are zero-modes in the single-instanton background, then there will be even more zero-modes in multi-instanton backgrounds. \footnote{In \cite{Dine:1988kq} it is argued that $(0,2)$-models with a discrete R-symmetry are also free of worldsheet instantons. Worldsheet instantons generate terms of the form $(27 \overline{27})^n$ in the effective Lagrangian, which lift the bundle moduli directions. The discrete R-symmetry prohibits such terms from being generated.} In the class of models where these zero modes are present for every holomorphic genus zero curve in $M$, every possible worldsheet instanton vanishes and cannot contribute to the spacetime superpotential. This renders the vacuum non-perturbatively stable. However, it turns out that finding explicit examples of the splitting of bundles is quite hard. Furthermore, it is likely these examples are extremely special, and unlikely to be representative of the larger moduli space of heterotic compactifications.

Fortunately, it turns that there is a rather large class of compactifications based on Calabi-Yau complete intersections, in which it is believed worldsheet instantons do not destabilise the vacuum. The instantons, while non-zero, actually cancel out in a rather remarkable fashion. These are conformal field theories built as the infrared fixed points of gauged linear sigma models, a study of which, we will now turn to.

\section{$(0,2)$ Linear Sigma Models}
\label{sect:glsm}
Gauged linear sigma models (GLSMs) have proven to be a versatile tool in exploring the moduli space of certain $(2,2)$ SCFTs. As the name suggests this is a two-dimensional gauge theory, consisting of a number of matter fields with flat kinetic terms
coupled to an abelian gauge group.\footnote{The gauge group can also be non-abelian, although these are associated with hypersurfaces in  Grassmannians and will not be relevant for the discussion in this review.} One way to think of the GLSM is that it builds conformal field theories that coincide with the infrared fixed points of NLSMs with a particular Calabi-Yau target space.  The UV theory (GLSM) and the IR theory (SCFT) are in principle completely different entities, so why is the GLSM useful? The answer lies in the existence of a number of quantities that are not renormalised under RG. These include F-terms of the low-energy effective field theory (unnormalised Yukawa couplings) and the singularity structure of the SCFT (points where string perturbation theory breaks down). Already this sheds a significant amount of light on the $d=4$ compactification, and as we shall see, these quantities are relatively straightforward to compute.

The parameters space of linear sigma models contain regions or `phases', in which the underlying SCFT is not necessarily associated with any non-linear sigma model. Instead, the conformal field theories are described by less geometric constructions, with the canonical example being Landau-Ginzburg theories. The GLSM has a natural generalisation to theories with $(0,2)$ supersymmetry, and will be used extensively in our discussion. As has been persuasively argued in~\cite{Silverstein:1995re,Basu:2003bq,Beasley:2003fx}, conformal field theories built as fixed points of the $(0,2)$ GLSM automatically evade the instanton corrections that destabilise the vacuum, and have proven to be a useful tool in constructing  $(0,2)$ SCFTs.

Despite a promising conceptual framework, little is understood about $(0,2)$ GLSMs, and there remain many open questions. Although the $(2,2)$ GLSM has been successfully used to understand questions such as the structure of the quantum moduli space of $(2,2)$ conformal field theories, and how quantum corrections modify low-energy observables such as Yukawa couplings, these remain open questions for general $(0,2)$-theories. Some of these questions have been answered for certain $(0,2)$ theories. These include $(0,2)$ theories attained as deformations of $(2,2)$ theories and exactly soluble models (analogous to Gepner models). The former class of theories have a geometric interpretation as a sigma model for a Calabi-Yau target-space equipped with a rank $3$ holomorphic vector bundle.  The (2,2) locus amounts to setting the holomorphic bundle to be the tangent bundle of the Calabi-Yau manifold, and the (0,2) deformations are holomorphic deformations of the tangent bundle.

In this section we aim to give some bare bones details of linear sigma models. As this material is well-known we will be brief, giving only the pertinent details for later sections and refer the reader to the original reference \cite{Witten:1993yc} for more details.

\subsection{The Bare Bone Basics: Field Content and Lagrangian}
The most convenient way to describe $(0,2)$ linear sigma models is in superspace with coordinates $x^\pm,\theta^+,\thetab^+$, superspace covariant derivatives $\cD_+$, $\cDb_+$, and supercharges $\cQ_+,\cQb_+$. For $(0,2)$ sigma models that are deformations of $(2,2)$ linear sigma models, the field content is most easily understood in terms of decomposing the $(2,2)$ multiplets into $(0,2)$ multiplets.

A $(2,2)$ gauged linear sigma model is an abelian gauge theory with dimensionful coupling $e_0$, linear kinetic terms and a gauge group $\U(1)^r$. The field content consists of $n$ chiral multiplets $\Phi^i_{(2,2)}$, $i=1,\ldots,n$, with gauge charges $Q_i^a$, and $r$ gauge multiplets $\Sigma_a$, $a=1,\ldots,r$. The action consists of the kinetic terms, a superpotential term and a twisted chiral superpotential term. The last term includes a possible Fayet-Iliopoulos term (see~\cite{Witten:1993yc} for more details). Away from the $(2,2)$ locus these pieces decompose into $(0,2)$ representations. The chiral multiplets decompose into $(0,2)$ multiplets as
\begin{equation}
\Phi_{(2,2)} \to (\Phi,~\Gamma),
\end{equation}
where $\Phi^i$ is a (0,2) chiral multiplet, and $\Gamma^i$ is a $(0,2)$ Fermi multiplet. The $(2,2)$ twisted chiral field-strength multiplet splits up as
\begin{equation}
\Sigma^{(2,2)} \to (\Sigma,~\Up),
\end{equation}
where $\Sigma$ is a (0,2) chiral superfield, and $\Up$ is $(0,2)$ Fermi multiplet.
Working in Wess-Zumino gauge, the field-strengths $\Upsilon$ have the superspace expansion
\begin{eqnarray}
%V_{a,-} & = & v_{a,-} - 2i \theta^+ \lambdab_{a,-} - 2i \thetab^+ \lambda_{a,-} + 2 \theta^+ \thetab^+ D, \nonumber\\
\Upsilon %& = & i \cDb_+ V_{a,-} + \theta^+ \p_- v_{a,+} \nonumber\\
~&=& -2 (\lambda_{a,-} - i \theta^+(D -i f_{a,01}) -i \theta^+\thetab^+ \p_+ \lambda_{-,a} ).
\end{eqnarray}
The bosonic multiplets  obey a chirality constraint $\cDb_+ \Phi = \cDb_+ \Sigma = 0$ and  have an expansion involving gauge-covariant derivatives $\nabla$:
\begin{eqnarray}
\Phi &=& \phi + \sqrt{2} \theta^+ \psi_+ -i \theta^+\thetab^+ \nabla_+ \phi, \nonumber\\
\Sigma_a & = & \sigma_a +\sqrt{2} \theta^+\lambda_{a,+} - i \theta^+\thetab^+ \p_+\sigma_a.
\end{eqnarray}
The fermionic matter multiplets $\Gamma$ are the most interesting new structures to emerge from
the (2,2)$\to$(0,2) reduction.  These fields are not chiral, but rather satisfy
\begin{equation}
\label{eq:GE}
\cDb_+\Gamma^A = \sqrt{2} E^A(\Phi,\Sigma),
\end{equation}
where on the (2,2) locus the index $A=i=1,\ldots,n$ and the functions $E^i$ are given by
\begin{equation}
\label{eq:22E}
E^i = i \sqrt{2} \sum_a Q_i^a \Phi^i \Sigma_a.
\end{equation}
Off the $(2,2)$ locus the $E^i$ are in general holomorphic functions of the $\Phi^i$ and $\Sigma_a$ multiplets.
The explicit superspace expansion is given by
\begin{eqnarray}
\Gamma^i &=& \gamma_-^i - \sqrt{2} \theta^+ G^i - i \theta^+\thetab^+ \nabla_+\gamma_-^i - \sqrt{2}\thetab^+ E^i(\Phi,\Sigma) \nonumber\\
~&= & \gamma_-^i -\sqrt{2} \theta^+ G^i -\sqrt{2}\thetab^+ E^i(\phi,\sigma)\nonumber\\
~&~&~ - i\theta^+\thetab^+\left[\nabla_+\gamma_-^i + 2i E^i_{~,j} \psi_+^j + 2 i E^i_{~,a} \lambda_{a,+} \right].
\end{eqnarray}
For $(0,2)$ theories not attainable as deformations of $(2,2)$ models, the number of fermi multiplets does not necessarily equal the number of chiral multiplets, reflected by the use of the index $A=1,\ldots,N$ and $i=1,\ldots,n$. For the most part of the next section we will set $A=i$ and $N=n$, though in section~\ref{s:lg} we will relax this.

The classical Lagrangian consists of the standard flat kinetic terms
\bea
\cL_{{\rm KE}} &=& \int d\theta^+ d \thetab^+ ~\ff{1}{8e_0^2}\Upsilonb_a \Upsilon_a + \ff{i}{2e_0^2}\Sigmab_a \p_- \Sigma_a +  \ff{i}{2}\Phib^i (\p_- + i Q_i^a V_{a,-} ) \Phi^i + \ff{1}{2}\Gammab^A \Gamma^A,\label{ke1}
\eea
together with a set of superpotential couplings given by
\begin{eqnarray}
\label{eq:jdef}
\cL_{\cJ} &=& -\ff{1}{\sqrt{2}} \int d\theta^+~ \Gamma^I \cJ_I(\Phi)|_{\thetab^+=0} + ~\text{h.c.}.
\end{eqnarray}
Not all of the $E$ and $J$ couplings are independent however, and are related by a supersymmetry constraint. This ensures the Lagrangian $\cL_J$ preserves $(0,2)$ supersymmetry:
\begin{equation}
\label{eq:02constraint}
\sum_A E^A \cJ_A = 0.
\end{equation}
On the (2,2) locus, $i=A$, $J_i = P_{,i}= \del P / \del \Phi^i$ and the constraint reduces to
\begin{equation}
\Phi^0 \Sigma_a \left[Q_0^a P + \sum_i Q_i^a P_{,i} \right] = 0,
\end{equation}
where the equality follows from the quasi-homogeneity properties of $P$ implied by gauge invariance.
Clearly, this is not the only way to satisfy the constraint.  Replacing the $P_{,i}$ with polynomials $J_A$ of
same charge, and choosing more general $E^A$ as we did in the $V$-model, we will find a theory with
(0,2) supersymmetry if
\begin{equation}
E^0 P + \Phi^0 E^AJ_A = 0.
\end{equation}
Finally, there is a twisted superpotential coupling realising a FI term
\begin{equation}
\label{eq:FI}
\cL_{\text{F-I}} = \ff{1}{4}\int d\theta^+ \sum_{a=1}^{r}\tau^a \Upsilon_a + \text{h.c.}.
\end{equation}
where $\tau^a$ are the complexified Fayet-Iliopoulos parameters: $\tau^a = i\rho^a + \theta^a/2\pi.$
The action has an classical symmetry $\GU(1)_L \times \GU(1)_R$ with charges display in table~\ref{table:U1vmodel}.
\begin{table}[t]
\begin{center}
\begin{tabular}{|c|c|c|c|c|c|}
\hline
$~			$&$\theta^+ 	$&$\Phi^i		$&$\Gamma^i	$&$\Sigma_a	 $&$\Upsilon_a 	 $\\ \hline
$\GU(1)_R	$&$1			$&$0			$&$0			$&$1			 $&$1			 $\\ \hline
$\GU(1)_L		$&$0			$&$0			$&$-1		$&$-1		 $&$0			 $\\ \hline
\end{tabular}
\end{center}
\caption{The $\GU(1)_L\times\GU(1)_R$ symmetry charges for the $V$-model.}
\label{table:U1vmodel}
\end{table}
On the (2,2) locus these are just the classical left-moving and right-moving R-symmetries. In the heterotic string the $\U(1)_L$ is a global symmetry used to defined the GSO projection as discussed in section \ref{symm}.

\subsection{Vacuum Structure and Phases}
\label{ss:phases}
The phase structure of the $(0,2)$ GLSM is studied in the same way as the $(2,2)$ model,  though one finds a richer structure. To illustrate the features we will use in this review, let us focus on the simple example of the quintic. In that case there are five chiral multiplets $\Phi^i$ of gauge charge $+1$, a single chiral multiplet $\Phi^0$ of charge $-5$ and a corresponding number of fermi multiplets $\Gamma^i,\Gamma^0$. The superpotential terms $\cJ^i = \phi^0 J_i(\Phi^i)$ and $\cJ^0 = P(\phi^i)$, where the vanishing of $P$ defines the hypersurface, while the holomorphic functions $J_i$ are tied up in defining the bundle on the quintic (on the $(2,2)$ locus $J_i = \del W / \del \phi^i$ with $W$ being the $(2,2)$ superpotential).

Integrating out the superspace coordinates in \C{ke1}-\C{eq:FI}, one finds a bosonic potential given by
\be
U \simeq |P|^2 + |\phi^0|^2 |J_i|^2 + \frac{e_0^2}{2} \left(\sum Q_i |\phi^i|^2 + Q_0|\phi^0|^2 - r\right)^2 +  E^A \Eb^A
\ee
and Yukawa interactions
\be
\cL_{yuk} \sim \gammab_-^A \psi_+^j \frac{\del E^A}{\del \phi_j}   +   \lambda_{+} \gammab_-^A \frac{\del E^A}{\del \sigma} +
\gamma_-^i \psi_+^j \frac{\del J_i}{\del \phi^j} + \gamma_-^i \psi_+^0 \frac{\del J_i}{\del \phi^0} + {\rm h.c.}
\ee
where $A=i=1,\ldots,n$. It turns out that for any hypersurface in a projective space we can, by a choice of field redefinitions, pick
\be
E^i = \phi^i \sigma, \quad E^0 = -5\phi^0\sigma
\ee
The {\it Calabi-Yau phase} (geometric phase) corresponds to $r\gg 0$, in which semi-classical analysis is a good approximation. The ground state is given by $\phi^0=0$ and $\sum Q_i |\phi^i|^2 = r$, which after quotienting by the $\U(1)$, gives a projective toric variety. We also have to set the superpotential $P = 0$, and this defines a hypersurface in the toric variety.

The massless spectrum of the worldsheet fields is easy to work out, and exhibits how the linear sigma model describes the bundle $\cF$. The right-moving fermion $\psi_+^0$ gets a mass together with one linear combination of the left-moving fermions $\gamma_-^i$ via the Yukawa couplings. It is not hard to see that the remaining left-moving fermions $\gamma_-^i$ are massless and transform as sections of the bundle $\cF$ defined by
 \begin{equation}
\xymatrix{ 0\ar[r] &\cO^{r}|_M \ar[r]^-{E} &\oplus_i\cO(D_i)|_{M}  \ar[r]^-{J} & \cO(\sum_i D_i)|_{M}\ar[r] & 0 },\label{sequence}
\end{equation}
Likewise, the left-moving fermion $\gamma_-^0$ gets a mass together with a linear combination of the right-moving fermions $\psi_+^i$ via the Yukawa coupling $\gamma_-^0 \psi_+^i $. There is also the Yukawa coupling $\sum_i \del_\sigma E^i \lambda \psi_+^i \bar\phi_i$ which gives a mass to another one of the fermions $\psi_+^i$. The remaining fermions $\psi_+^i$ remain massless and transform as local sections of the tangent bundle $TM$. Thus, we see the right-moving fermions are sections of the tangent bundle, while the left-moving fermions are sections of the gauge bundle in agreement with our intuition from the non-linear sigma model constructions

The {\it Landau-Ginzburg phase} happens at small radius corresponding to $r \ll 0$. The minimum of $U$ is at $|\phi^0|^2 = |r|/5$, $\phi^i = 0$. The gauge symmetry is spontaneously broken, and as $\Phi^0$ has charge $-5$ it is broken to the discrete subgroup $\Z_{5}$.  The $\phi^0$ field becomes massive being integrated out of the low-energy theory. The fermion $\psi^0$ is also massive via a Yukawa coupling. This leaves one with a $\Z_{5}$ Landau-Ginzburg orbifold described by the superpotential
\be
S_{LG} = \int d^2\sigma d\theta^+\, \Gamma^0 P(\Phi) + \Gamma^i J_i(\Phi).
\label{eq:lgdef}
\ee
The point $r=\theta=0$ corresponds to a singularity in the low-energy theory. A non-compact direction in field space is opened up. Nonetheless, the singularity is codimension one, and one can show the spectrum of low-lying states is continuous as one varies from $r\gg0$ to $r\ll 0$. The phase picture becomes more intricate when we consider more gauge groups and complete intersections, though is straightforward to work out (see \cite{Distler:1993mk}).
Although we have restricted to a simple example it is straightforward to extend this to more intricate examples, involving higher rank gauge groups. For example, for a suitable choice of $\rho^a$, the real part of $\tau_a$, the classical moduli space of the gauge theory is
\begin{equation}
\cM_{0} (r) = \left\{ D_a = {\textstyle\sum_i} Q_i^a |\phi^i|^2 -\rho^a = 0 \right\}/ [\GU(1)^r],
\end{equation}
which is a symplectic quotient presentation of a toric variety. A more detailed discussion is given in say \cite{Morrison:1994fr}.

\subsection{The Parameter Space of $(0,2)$ Linear Sigma Models}
\label{ss:param}
The linear sigma model with $(2,2)$ supersymmetry has a parameter space that does not capture the entire moduli space of the string compactification. It only captures the ``toric'' subset of K\"ahler moduli, and complex structure deformations that are representable as polynomial deformations. What bundle deformations of the standard embedding do linear sigma models capture? Said differently, how many $(0,2)$ deformations of $(2,2)$ models are there? This was answered in recent work\cite{Kreuzer:2010ph}, to which we will now summarise. A warning to the reader: this subsection contains an amount of toric geometry. Reviews on toric geometry for physicists may be found in a number of places e.g. \cite{Kreuzer:2010ph,Morrison:1994fr,Aspinwall:1993nu}.

For example, let us study the linear sigma model that describes a hypersurface in a toric variety $V$. This is termed the $M$-model, and will be studied in more detail below. The field content, described in the previous subsection above, has $n+1$ Fermi multiplets obeying the constraint \C{eq:GE} which introduces the E-parameters. It is straightforward to compute the number of these parameters being given by
\begin{equation}
\# E  = r (1+\dim\Autt V ).
\end{equation}
where $\dim\Autt V$ is related to the automorphism group of the toric variety $V$ via the exact sequence\footnote{As we are interested in deformations of the tangent bundle, we consider only the connected component of the Automorphism group of $V$.}
\begin{equation}
\xymatrix{1 \ar[r] & G \ar[r] & \Autt V \ar[r]& \Aut V \ar[r] & 1,}
\end{equation}
where $G=\U(1)^r$ for us, so that essentially $\dim \Autt V = \dim \Aut V + r$. The $M$-model has a number of superpotential couplings given in \C{eq:jdef}. The number of parameters in the $(0,2)$ superpotential is counted by the number of parameters in the $J_i$ functions. This turns out to be
\begin{equation}
\label{eq:Jnumgen}
\# J = n \ell(\Delta) - \sum_{\rho\in\Delta^\circ}  \sum_{m\in\Delta} \delta (\la m, \rho \ra + 1)
\end{equation}
where $l(\Delta)$ is the number of lattice points interior to the Newton polytope $\Delta$ defining the hypersurface $P=0\subset(\CC^*)^d$. This corresponds to the number of independent monomials making up the generic polynomial $P$. The second term is trickier to explain. $\Delta^\circ$ is the polytope dual to $\Delta$, with lattice points along the edges of $\Delta^\circ$ corresponding to the coordinates of the sigma model (there are $n$ of them in total). As these are dual polytopes, elements of the polytopes have a natural inner product $\la~,~\ra$, and the delta function then implies the second term counts the number of points whose natural inner product is $-1$. We also need to take into account that not all of $E$ and $J$ parameters are independent, with some being related by the supersymmetry constraint \C{eq:02constraint}. This eliminates $rl(\Delta)$ parameters.

We need to take into account some of the parameters entering into the Lagrangian may be eliminated with field redefinitions that modify only the presumably irrelevant D-terms of the linear sigma model. For example, there are the rescalings of $\Phi$ and $\Gamma$ multiplets, as well as the $\GL(r,\CC)$ rotations of the $\Sigma_a$ multiplets. Field redefinitions of the $\Phi^i$ turn out to eliminate
\begin{equation}
\#\delta \Phi =  \dim\Autt V
\end{equation}
parameters. Similarly redefinitions of the fermionic multiplets eliminate
\begin{equation}
\#\delta\Gamma = \#\delta \Phi + \sum_{\varphi}\sum_{m\in\ri\varphi}\left[n-2- \sum_{\rho} \delta(\la m,\rho\ra)\right],
\end{equation}
where $\varphi$ are the facets (codimension 1) faces of the polytope $\Delta$ and $\ri\varphi$ denotes their relative interior. Note that when $V$ is a product of projective spaces $\#\delta\Gamma = \#\delta \Phi$. We must also take into account that field redefinitions that act as gauge or $\U(1)_L$ transformations do not act on the action.

We are almost ready to count the number of deformations counted by the linear sigma model. The remaining aspect we need to account for is a subtle issue of resolving singularities in the ambient toric variety. Generically, the toric variety will have a number of suitably mild singularities, say $w$ of them, that may be resolved in a fashion that preserves the properties of the hypersurface, in particular that it is smooth and Calabi-Yau (for a summary of this from the point of view of toric geometry see \cite{Kreuzer:2010ph}). For each resolution, the corresponding sigma model has an additional K\"ahler parameter and for the class of Calabi-Yau's we are interested in these singularities will not intersect the hypersurface. From the point of view of the linear sigma model, this means the corresponding $w$ K\"ahler parameters are irrelevant in the IR and do not correspond to real deformations of the SCFT. These $w$ parameters are to be discounted from the parameter count. Altogether then we have the following count of deformations
\begin{eqnarray}
N(M) & = & 1-w -(r-1)( r+\#P-\#\delta \Phi-2) + \#J - \#\delta\Gamma.
\end{eqnarray}
A special class of $M$-models are known as reflexively plain. Geometrically this occurs when the automorphism groups are as small as possible---the automorphisms of $M$ are those inherited form the $[\CC^*]^d$ reparameterisations of the algebraic torus of $V$. From the point of view of the linear sigma model this means there are no `off-diagonal' E-deformations: $E_a^i(\phi) = e_a^i \phi_i$ (for $a=1,\ldots,r$ with no sum on $i$) and the field redefinitions are simple rescalings\footnote{ Combinatorially, this means there are no interior points of facets of the Newton polytope $\Delta$ or its dual $\Delta^\circ$.}
$$\Phi_i \rightarrow u_i \Phi_i\quad\Gamma^A \rightarrow v_A\Gamma^A\quad({\rm no~sum}).$$
If we define a mirror hypersurface $M^\circ$ given by exchanging $\Delta$ with $\Delta^\circ$, then $M$ and $M^\circ$ give rise to a pair of linear sigma models with the same number of deformations $N(M) = N(M^\circ)$. This suggests some type of map between the two models, which we will explore in section~\ref{ss:mirror} below.

If a model is reflexively plain then there is a convenient parameterisation of the deformations that is manifestly invariant under field redefinitions. As the field redefinitions and E-parameters are diagonal, it is convenient to combine the $r$ $\Sigma_a$ multiplets into a vector $\bSigma$. Then, we can express the E-couplings as
\begin{align}
\label{eq:Edefsgen}
E^0 = \Phi_0 \bSigma\cdot \bdelta, \quad E^{i} =  \bSigma\cdot \bE^{i}(\Phi),
\end{align}
where $\bdelta$ is a vector of parameters. The polynomial defining the hypersurface $M\subset V$ is given by
\be
P(\phi) = \sum_{m\in\Delta} \alpha_m \mu_m,
\ee
where $\mu_m$ are the monomials in $\phi^i$ defined by the polytope $\Delta$. Similarly, homogeneity of $J_i(\phi)$ implies
\be
\Phi_i J_i = \sum_{m\in\Delta} j_{m i} \mu_m,\quad({\rm no~sum~on~}i)
\ee
Under a set of chiral field redefinitions, ones with $u_i\ne v_i$, the path integral measure transforms anomalously, resulting in an effective shift of the K\"ahler parameters $q_a$. Unlike $(2,2)$ theories, the parameters $q_a$ are ambiguous in $(0,2)$ theories.  One can then show that a set of invariant ``K\"ahler'' parameters under these field redefinitions is given by
\begin{equation}
\kappa_a  \equiv q_a \prod_i \left(\frac{j_{0i}}{\alpha_0}\right)^{Q^a_i}.
\end{equation}
Note that on the $(2,2)$ locus these reduce to the usual $q_a$. A set of invariant ``complex structure'' coordinates is given by
\begin{equation}
\kappah_{\ah} \equiv \prod_{m\neq 0} \left( \frac{\alpha_m}{\alpha_0} \right)^{\Qh^{\ah}_m}.
\end{equation}
We also define
\begin{equation}
\bgamma^i \equiv \frac{j_{0i}}{\alpha_0} {\bf e}^i, \quad\text{and}
\quad b_{mi} \equiv \frac{\alpha_0 j_{mi}}{\alpha_m j_{0i}}-1\quad\text{for}\quad m\neq 0.
\end{equation}
Assuming $\alpha_m\ne0$, the supersymmetry constraint $E\cdot J = 0$ implies
\begin{equation}
\bdelta = - \sum_i \bgamma^i, \quad  \sum_i b_{mi}\bgamma^i = 0 \quad \text{for} \quad m \neq 0.
\end{equation}
On the $(2,2)$ locus the matrix $b_{mi}$ has rank $d$, which determines the $\bgamma^i$ to be in the r-dimensional kernel of $b$, implying $\bgamma^i$ and $\bdelta$ are determined up to $\GL(r,\CC)$ field redefinitions. In fact, the only time $b$ is not rank $d$ is when the model is singular and the bundle jumps rank becoming a sheaf. The utility of this parameterisation will become clear when we come to discussing the singular locus and $(0,2)$ mirror map.

\subsection{Twisting $(2,2)$ Linear Sigma Models}
The (2,2) linear sigma model, like its non-linear cousin, admits topological twistings. This involves shifting the Lorentz generator by a linear combination of the $\U(1)_L$ and $\U(1)_R$ R-symmetries.  There are essentially two possibilities: twisting by a vectorial combination (A-twist) and twisting by an axial combination (B-twist). As we consider heterotic string compactifications the $(0,2)$ linear sigma models we consider still have a $\U(1)_L\times\U(1)_R$ symmetry, with the former now becoming a  global symmetry. This allows us to define the generalisations of these twists which we denote the A/2-twist and B/2-twist.

Just like $(2,2)$ models, this results in the spins of fields shifting---for example, all the fermions become either scalars or one-forms. Unlike $(2,2)$ models there is only one supercharge that becomes a scalar. Hence, the BRST cohomology is defined by the kernel of the scalar right-moving supercharge $Q$. This cohomology ring is infinite dimensional as opposed to the $(2,2)$ case, so it is not a priori obvious that many of the techniques, for example localisation, will carry over. However, the presence of the $\U(1)_L$ allows us to define a finite dimensional subring\cite{Adams:2003zy,Adams:2005tc} with which many of the localisation techniques hold. In particular, one looks for states that satisfy the usual chiral ring relation $h_{L,R} = \pm \half q_{L,R}$;\footnote{A lovely explanation of the chiral ring in $N=2$ SCFTs is given in \cite{Warner:1993zh}.} the ground ring is then shown to coincide with the usual chiral ring on the $(2,2)$ locus, and also exists when there is no $(2,2$) locus.

\subsection{A/2-Twisted $V$-model}
In the absence of a superpotential, the sigma model, called here the $V$-model, describes a target space given by the toric variety $V$. As this space is not Ricci-flat the sigma model  does not actually flow to a conformal field theory, instead flowing to a trivial fixed point. Nonetheless, it is possible to construct an A/2-twisted $V$-model and it may be shown to have a well-defined notion of a quantum cohomology ring. Furthermore it proves to be a useful warm up for the $M$-model where there is a non-trivial superpotential.

The A/2 twist of the (0,2) NLSM with toric target-space was considered in~\cite{Katz:2004nn}.  The point of view advocated in~\cite{Katz:2004nn} was to combine the familiar structure of (2,2) worldsheet instantons with the notion that in (0,2) theories the basic A/2 twisted observables (the $\sigma_a$ in our case) should correspond to classes in $H^1(V,\cF^\vee)$.  Classically (i.e. for constant maps), the computation of a correlator is reasonably clear: $\la \sigma_{a_1}\cdots \sigma_{a_d} \ra$ should yield a map
\begin{equation}
H^1(V,\cF^\vee) \times H^1(V,\cF^\vee) \times \cdots \times H^1(V,\cF^\vee) \to H^d(V, \wedge^d \cF^\vee) \simeq H^{d,d}(V)\simeq \CC.
\end{equation}
The second-to-last isomorphism automatically holds in theories with a (2,2) locus~\cite{Adams:2005tc}.

By using the universal instanton construction, the authors of~\cite{Katz:2004nn} described how to pull back the bundle (more generally, sheaf) $\cF$ to a sheaf on $\cM_n$ and in principle compute the induced sheaf cohomology on the instanton moduli space.   As usual in NLSM computations, these results required some choice of compactification of the instanton moduli space.  In the case when $V$ is a toric variety, the GLSM naturally provides such a compactification.  The ideas in~\cite{Katz:2004nn} were refined and developed in~\cite{Guffin:2007mp}, culminating in a general method for computing the A/2 correlators in the $V$-model. The result should be thought of as a quantum deformation of the sheaf cohomology on $H^\ast (V,\wedge^k \cF^\vee)$.

While the method of~\cite{Katz:2004nn,Guffin:2007mp} is well-motivated and leads to sensible results, it is desirable to compute correlation functions directly in the sigma model. This was worked out in detail in \cite{McOrist:2008ji,McOrist:2007kp} and led to a number of interesting conclusions. Firstly, it directly showed the sheaf cohomology computations of Katz-Sharpe did indeed coincide with the correlators computed by the linear sigma model. Secondly, it gave a direct way of computing the quantum cohomology relations (as opposed to computing correlators and extracting the relations indirectly). Finally, by studying the linear sigma model action and zero mode structure in an instanton background, it gave a natural generalisation of toric intersection theory to $(0,2)$ sheaf cohomology. Let us outline some of these results in a bit more detail. In the following sections we will always assume the sigma models have a $(2,2)$ locus, and therefore not distinguish the $A$ and $i$ indices.

\subsubsection{Coulomb Branch}
One approach for deriving the $(0,2)$ quantum cohomology relations is when the linear sigma model has a Coulomb branch\cite{McOrist:2007kp}. When $V$ is Fano, then there always exists a region in the K\"ahler parameter space where supersymmetry is broken classically. In fact, this is not true in the quantum theory: quantum effects restore supersymmetry resulting in a number of discrete Coulomb vacua. In this phase the $\sigma$ fields obtain large VeVs, the $\Phi^i,\Gamma^i$ matter multiplets get massive, and the dynamics of the $\Sigma_a$ multiplets are determined by an effective twisted superpotential given by
\begin{equation}
\label{eq:Coulomb}
\cL_{\text{eff}} =   \int d \theta^+ \sum_{a=1}^{r}\Upsilon_a \Jt_a(\Sigma)|_{\thetab^+=0} + \text{h.c.},
\end{equation}
with
\begin{equation}
\label{eq:J}
\Jt_a  =  \log \left[ q_a^{-1} \prod_\alpha \det M_{(\alpha)}^{Q_{(\alpha)}^a} \right],
\end{equation}
where the $M_{(\alpha)}$ are given by
\begin{equation}
\label{eq:Edef}
\cDb_+ \Gamma_{(\alpha)} = 2 i M_{(\alpha)} \Phi_{(\alpha)},~~~M_{(\alpha)} = \sum_{a=1}^{r} \Sigma_a E^a_{(\alpha)},
\end{equation}
where $\Phi_{(\alpha)}$ is a vector of fields of the same gauge charge of length $k_{(\alpha)}$, and $M_{(\alpha)}$ is a $k_{(\alpha)}\times k_{(\alpha)}$ matrix mixing these fields. This effective superpotential encodes the quantum cohomology relations of the A/2-twisted $V$-model, and localisation techniques applied in the non-geometric phase yield the correlators in the $V$-model. These results are all natural generalisations of the $(2,2)$ A-twisted model discussed in \cite{Witten:1993yc,Morrison:1994fr,Melnikov:2006kb}. Indeed, the effective potential allows us to read off the quantum cohomology relations directly:
\begin{equation}
\label{eq:02QC}
\la \sigma_{a_1} \cdots \sigma_{a_k} \prod_{\alpha|Q^a_{(\alpha)}>0} \det M_{(\alpha)}^{Q^a_{(\alpha)}} \ra =
q_a \la \sigma_{a_1} \cdots \sigma_{a_k} \prod_{\alpha|Q^a_{(\alpha)}<0} \det M_{(\alpha)}^{-Q^a_{(\alpha)}} \ra~~\text{for all}~a.
\end{equation}
As on the (2,2) locus, it is easy to extend this description to an explicit formula for the genus zero A/2-twisted correlators.  A simple generalisation
of the localisation formulae in half-twisted Landau-Ginzburg models yields the correlators as a sum over the common zeroes of the $\Jt_a(\sigma)$:
\begin{equation}
\label{eq:Coulcorr}
\la \sigma_{a_1} \cdots \sigma_{a_k} \ra = \sum_{\sigma| \Jt = 0}\sigma_{a_1}\cdots\sigma_{a_k} \left[ \det_{a,b} \Jt_{a,b} \prod_\alpha \det M_{(\alpha)}
\right]^{-1}.
\end{equation}
As expected, the correlators are position-independent, given by meromorphic functions of the $q_a$ and the
E-deformations, and satisfy the quantum cohomology relations.  When applied to the example
of $V\simeq \P^1\times \P^1$, the results are in agreement with the computations of~\cite{Guffin:2007mp}.

\subsubsection{Gauge Instantons}
An alternative approach to understanding the correlation functions and quantum cohomology is via the geometric phase, in which case the D-terms have classical solutions given by gauge instantons. In that case the linear sigma model reduces to a non-linear sigma model with a geometric target space. In that case the relevant BPS configurations are given by
\begin{equation}
\label{eq:02Vloc}
\p_{\zb} \sigma_a = 0,~~~{ E^{ai} \sigma_a = 0}~~(\text{no sum on}~i),~~~\nabla_{\zb} \phi^i = 0,~~~D_a + f_a = 0,
\end{equation}
Comparing these conditions to those of the topological theory at the (2,2) locus, we see that as long as
$E^{ai}(\phi) $ has rank $r$ for all $\phi$ outside the exceptional set (this will be true for small E-deformations), the only solution to the
first two conditions is $\sigma_a=0$, and the last two are solved by gauge instanton configurations whose topological class is labelled by instanton numbers
\begin{equation}
n_a = -\ff{1}{2\pi} \int f_a.
\end{equation}
Thus we find that just as on the $(2,2)$ locus, the moduli space of BPS configurations is given by the moduli space of gauge instantons $\cM_n$. We can compute the quantum cohomology of the $(a,c)$-ring by evaluating the gauge instanton configurations sector-by-sector. For $(2,2)$ linear sigma models, this is explained in detail in \cite{Morrison:1994fr}. It essentially involves reducing the calculation to an integral over the zero modes of the instanton configuration. The moduli space of which is described is a compact toric variety $\cM_n$. Each of the chiral ring elements $\sigma_a$ lift to a cohomology class  $\eta_a \in H^2(\cM_n,\Z)$. The correlator evaluated in the $n$th gauge instanton sector $$\la\sigma_{a_1}(x_1)\cdots\sigma_{a_k}(x_k)\ra$$ is then given by doing a corresponding intersection computation on $\cM_n$, denoted by $\#(\ldots)$:
\begin{equation}
\la\sigma_{a_1}(x_1)\cdots\sigma_{a_k}(x_k)\ra \leftrightarrow \#(\eta_{a_1}\cdots \eta_{a_k} )_{\cM_n}.
\end{equation}
As $M_n$ is toric, there are well-defined rules for doing this computation. The final result is then given by a sum over gauge instantons:
\begin{equation}
\label{eq:V22corr}
\la \sigma_{a_1}\cdots\sigma_{a_k} \ra = \sum_{n\in \cK^\vee}\#(\eta_{a_1}\cdots\eta_{a_k} \chi_{n} )_{\cM_n}\prod_{a=1}^{r} q_a^{n_a},
\end{equation}
where
$q_a = e^{2\pi i \tau^a}$ and $\chi_n$ is the Euler class of a certain obstruction bundle.

It is shown in \cite{McOrist:2008ji} that this generalises in a natural way to $(0,2)$ theories once one has developed a generalisation of toric intersection theory to sheaf cohomology on $H^*(V,\cF^\vee)$. This is an alternative approach to that developed in~\cite{Katz:2004nn}, relying more on toric type computational techniques that the linear sigma model naturally generates. We will not describe the toric intersection theory in detail here, instead referring the reader to the original reference\cite{McOrist:2008ji}.

\subsection{A/2-Twisted M-Model and $\brep{27}^3$ Yukawas}
With a non-trivial superpotential turned on, the linear sigma model now describes a hypersurface $M\subset V$. In order to consistently construct the model we need to add an additional chiral multiplet $\Phi^0$ with gauge charge $Q_0^a = -\sum_i Q_i^a$. The superpotential couplings are then
$$
\cL_{\cJ} = \int d\theta^+ \left[ \Gamma^0 P(\Phi^1,\cdots,\Phi^n) + \sum_{i=1}^n \Gamma^i \Phi^0 J_i\right] + \text{h.c.},
$$
where on the $(2,2)$ locus, $J_i = \p P/\p\Phi^i.$

The $M$-model has two types of (0,2) parameters:  the E-parameters familiar from the $V$-model, and the J-parameters appearing in the superpotential couplings above.
The two sets are not independent but must satisfy the (0,2) SUSY constraint. The geometric structure encoded by the $E$ and $J$ is a choice of bundle $\cF$ on the Calabi-Yau hypersurface $M \subset V$.
$\cF$ is a deformation of $T_M$, whose sections are described as the cohomology of the sequence
\begin{equation}
\xymatrix{ 0\ar[r] &\cO^{r}|_M \ar[r]^-{E} &\oplus_i\cO(D_i)|_{M}  \ar[r]^-{J} & \cO(\sum_i D_i)|_{M}\ar[r] & 0 },
\end{equation}
$\cF = \ker J / {\rm im} E$.
Physically, this sequence arises in the geometric phase of the GLSM as a description of the fermions in the
low energy NLSM~\cite{Witten:1993yc,Distler:1995bc}.

The A/2-twist of the $M$-model proceeds in an analogous fashion to the V-model. The supercharge $Q_T=\overline{\cQ}_+$ becomes a nilpotent scalar whose cohomology is represented by the $\sigma_a$ fields and correspond to elements of $H^{1,1}(M)$ being pull-backs of elements of $H^{1,1}(V)$. All the anti-holomorphic components are $Q_T$-trivial, and under some reasonable assumptions, correlators do not depend on them.
The moduli space of BPS configurations is that of the $V$-model \C{eq:02Vloc} together with
\begin{equation}
\phi^0 J_i = 0~~\text{for}~i>0,~~~\text{and}~~~P(\phi) = 0.
\end{equation}
This defines a hypersurface $M\subset V$ together with $\phi^0=0$. This defines a moduli space $M_{n;P}\subset M_{n}$ which while still compact, is difficult to describe. Most importantly it is no longer toric and the $V$-model methods do not directly apply. This is reminiscent of the $(2,2)$ linear sigma model analysis, which as argued in \cite{Morrison:1994fr}, one can relate $M$-model correlators to $V$-model correlators via the {\em quantum restriction formula}. In \cite{McOrist:2008ji}, the quantum restriction formula is generalised to $(0,2)$ linear sigma models given by deformations of $(2,2)$ theories. In particular an $M$-model correlator (denoted by $\lad \ldots \rad_M$) is related to $V$-model correlator via
\begin{equation}
\label{eq:02restrict}
\lad \sigma_{a_1} \cdots \sigma_{a_d} \rad_M = \la  \sigma_{a_1} \cdots \sigma_{a_d}  \frac{-M_0}{1-M_0} \ra_V,
\end{equation}
where $M_0$ is defined as
\begin{equation}
\cDb_+ \Gamma^0 = 2i M_{(0)} \Phi^0.
\end{equation}
On the (2,2) locus $M_{(0)} = -\sum_i Q_i^a \sigma_a$ and corresponds to the anti-canonical divisor on $V$. The left-hand side of \C{eq:02restrict} is the $M$-model correlator, while the right hand side is a $V$-model correlator. One can then compute the correlator using either a sum over gauge instantons or the Coulomb branch technique.

The correlator computed in \C{eq:02restrict} involves $(a,c)$-ring elements $\sigma_a$ and are related to the $\brep{27}^3$ Yukawa couplings of the physical theory. Although this is part of the story in computing $\brep{27}^3$ Yukawas, the computation is still not complete. Essentially, what \C{eq:02restrict} determines are the F-terms of the low-energy effective field theory; one still needs to compute the D-terms, which is equivalent to normalising the fields.

Although in the above we have essentially restricted to Calabi-Yau's built as hypersurfaces, these result generalise rather easily to complete intersections. Some of this has been developed in \cite{McOrist:2008ji}.
\subsubsection{Singular Locus}
The quantum restriction formula gives a simple way to compute the A/2-twisted $M$-model correlators.  From these
we may extract the quantum cohomology relations and determine the locus in the $q_a, M_{(\alpha)}$ parameter space
where the correlators have poles.  As on the (2,2) locus, these singularities should signal a singularity in the (0,2)
SCFT.  As in type II theories, we expect that here world-sheet perturbation theory breaks down and non-perturbative effects
are necessary to resolve the SCFT singularity.  These effects are not well understood in the heterotic string, and a
parametrisation of the singular locus in parameter space is an important step in studying this phenomenon.

In (2,2) theories it is well-known that the singular locus of the GLSM may be determined without computing a single correlator. The basic tool used is the effective potential governing the $\Sigma_a$ multiplets at large $\sigma_a$ VeVs.  This potential is easily obtained by integrating out the $\Phi^i$ multiplets at one loop.  We have already discussed how a similar potential may be computed off the (2,2) locus---a similar potential may be used to study the singular locus of the theory. After integrating out all the massive multiplets, in a similar fashion to what is done in $(2,2)$ theories, one finds a twisted superpotential similar to that in eqn.~(\ref{eq:Coulomb}):
\begin{eqnarray}
\label{eq:MJ}
\prod_\alpha \det M_{(\alpha)}(z)^{Q^1_{\alpha} }& = & (-\Delta)^{\Delta} q_1, \nonumber\\
\prod_\alpha \det M_{(\alpha)}(z)^{Q^a_{\alpha}} & = & q_a~~~\text{for}~ a>1,
\end{eqnarray}
where $\Delta = Q_0^1$ (we are in a basis where $Q_0^a = 0$ for $a>1$). These are $r=n-d$ equations for $r-1$ variables $z_a = \sigma_a/\sigma_1$. Generically, these equations are over-determined and do not have a solution. However, there are non-generic loci where these equations do have a solution. This indicates a singularity in the theory, since the $\sigma_a$ are fixed only up to an overall scale. This leads to a non-compact direction in field space, leading to a divergence in the $\sigma$ correlators. By studying the correlators computed by the quantum restriction formula \C{eq:02restrict} one can check they diverge precisely at the points where the equations \C{eq:MJ} are satisfied.

The singular locus takes an elegant form when $M$ is reflexively plain. Using the parameterisation given at the end of section~\ref{ss:param} we find
\begin{equation}
\prod_\rho \left[ \frac{\bsigma\cdot {\bf e}^\rho}{\bsigma\cdot\bdelta}\right]^{Q^a_\rho} = q_a,
\end{equation}
have a solution for some $\bsigma \neq 0$.  This is nicely rewritten in terms of our invariant coordinates as
\begin{equation}
\label{eq:A2locus}
\prod_\rho \left[ \frac{\bsigma\cdot \bgamma^\rho}{\bsigma\cdot\bdelta}\right]^{Q^a_\rho} = \kappa_a.
\end{equation}
The interpretation of this singularity in terms of the bundle is that the rank of $\cF$ increases at the singular point, implying that $\cF$ is no longer a bundle, but a sheaf. We also note that a classical singular bundle can lead to a well-defined theory away from large radius.

\subsection{B/2-twisted $M$-model and $\rep{27}^3$ Yukawas}
The $M$-model also admits a B/2-twist via the axial combination of $\U(1)_L\times\U(1)_R$. This twisting leads to the same $Q_T$ as the A/2-twisted theory: namely $\overline{\cQ}_+$ becomes a nil-potent scalar operator. It's unsurprising therefore that the half-twisted theory localises onto $M_{n;P}$---the same locus as the A/2-twisted theory. However, the theories are quite distinct: the twistings of the fields in the linear sigma model are different, resulting in a different set of local observables and different non-vanishing correlators. Just as for the B-twisted model, the gauge invariant observables are of the form $O_\alpha = \phi^0 f_\alpha(\phi)$ where $f_\alpha(\phi)$ is a polynomial in the $\phi^i$ fields. On the $(2,2)$ locus these are the gauge invariant monomials that appear in the superpotential, and correspond to complex structure moduli. Off the $(2,2)$ locus these operators remain perfectly well-defined, and form a basis for the $B/2$-twisted cohomology, being analogous to the $(c,c)$ chiral ring.

On the $(2,2)$ locus, the B-model is independent of the K\"ahler parameters $q_a$ and hence of quantum corrections. For $(0,2)$ theories the story is not as clean. What one can show is that there exist a ``large'' class of models that are independent of $q_a$ and hence quantum corrections. In fact, one can do better. If these models have a Landau-Ginzburg phases then the theory is actually independent of the E-parameters. This implies a nice decoupling argument: A/2-twisted correlators depend only on $q_a$ and E-parameters; B/2-twisted models (satisfying certain conditions) depend only the complex structure moduli and J-parameters. This decoupling is important to any generalisation of mirror symmetry to $(0,2)$ theories, a point with which we will return to momentarily.

To determine which model are independent of quantum corrections (i.e. the $q_a$) one needs to study the fermion zero modes in a given instanton background. That is, suppose we are computing a correlator of observables in a gauge instanton background with instanton number $n$. Then,
$$ \la \cO_1\ldots \cO_s\ra_n=0$$ unless
\begin{enumerate}
\item $s = d-1$ (ghost number selection rule)
\item $d_0  = \sum_{i=1}^n Q_i^a n_a = 0$ which follows from the structure of the $\phi^0$ zero modes. If we are in a product of projective spaces, this suffices to show all gauge instanton corrections vanish.
\item $I_+ = I_-$ and $\sum_{i\in I_{<-1}} (-d_i-1) = \sum_{i\in I_{>1}} (d_i - 1)$ where $d_i = \sum_{a=1}^r Q_i^a n_a$ and we have defined a number of sets:
\begin{eqnarray}
I_- &  = & \left\{ i \in I | d_i < 0 \right\}, \nonumber\\
I_0 &  = & \left\{ i \in I | d_i = 0 \right\}, \nonumber\\
I_+ &  = & \left\{ i \in I | d_i > 0 \right\}.
 \end{eqnarray}
 with similar definitions applying for $I_{<-1}$ and $I_{>1}$.
 \item $I_0 \ge r-1$ which follows by studying the gaugino $\lambda_a$ zero modes.
\end{enumerate}
Unfortunately, these elegant conditions are not sufficient to rule out non-trivial instanton contributions in all generality.  Consider the two-parameter $M$-model with charges
\begin{equation}
Q = \begin{pmatrix}-4 & 1 & 1 & 0 & 0 & 1 & 1 \\ 0&0 & 0 & 1 & 1 & -1 & -1 \end{pmatrix}.
\end{equation}
with hypersurface defined by the vanishing of
\begin{equation}
P = \phi_1^4 + \phi_2^4 + (\phi_3^4 + \phi_4^4+ \phi_3^2 \phi_4^2) \phi_5^4 + (\phi_3^4 +\phi_4^4) \phi_6^4.
\end{equation}
It is easy to see that $P= 0$ defines a smooth hypersurface in $V$, and that all the selection rules are satisfied. Consequently, the zero mode analysis is not enough to rule out quantum corrections in this model. What is one to make of this?

An interesting perspective on this question was found in~\cite{Sharpe:2006qd} for $(2,2)$ non-linear sigma models. In that case, one can either construct the B-twist or B/2-twist. In the former, the theory is manifestly independent of quantum corrections; in the latter it is not at all obvious and in particular, it seems possible that non-trivial gauge instantons may contribute to correlators. What \cite{Sharpe:2006qd} show is that when the selection rules permit gauge instantons to contribute, the integral of the bosonic zero modes amounts to integrating an exact form over an instanton moduli space.  If there is to be any non-zero contribution, it must comes from the boundary of the moduli space. Thus if one is working with compact moduli spaces this contribution is zero, and in particular, the linear sigma model moduli spaces are nicely compact. However, \cite{Sharpe:2006qd} worked with non-compact toric Calabi-Yau's---it is not clear how these results generalise to compact Calabi-Yau's in which the moduli space $M_{n;P}\subset M_n$ is no longer toric. But it is very tempting to believe that this analysis should carry over for Calabi-Yau hypersurfaces. Thus, it remains an interesting open question: are there really quantum corrections in such B/2-twisted linear sigma models?

What if we are working with a model in which there are no quantum corrections? In that case the theory is independent of the $q_a$ and answers do not depend where in the K\"ahler moduli space the computation is done. If the linear sigma model admits a Landau-Ginzburg description,\footnote{See \cite{Clarke:2010ep} for recent work in deriving conditions for the existence of affine Landau-Ginzburg models.} then one can show that correlators reduce to Landau-Ginzburg orbifold computations familiar from studies of $(2,2)$ models. These correlators are computed using the standard residue formulae~\cite{Vafa:1990mu,Vafa:1989xc,Melnikov:2007xi} and are related to the $\rep(27)^3$ Yukawa couplings (up to the question of computing the relevant D-terms, thereby normalising the matter fields in the spacetime effective field theory). A second interesting fact is in these class of models E-parameters decouple from physical observables.  A clue to seeing how this works is to realise that deep inside the Landau-Ginzburg phase the $\Phi^0,\Gamma^0$ and $\Sigma_a$ multiplets become massive, thereby decoupling. Hence, the resulting light fermion multiplets obey the constraint $\overline{\cD}_+ \Gamma^i=0$, and independent of variations of the E-parameters. We have not been overly careful in coming to this conclusion and the interested reader is urged to look at the original reference for more details\cite{McOrist:2008ji}.

\subsubsection{Singular Locus}
The singular locus of the B/2-twisted model, as in the B-twisted M-Model, comes from the $\Phi^0$ multiplet becoming light. This happens if and only if there exists a point $p\in M$ such that $J_i(p)=0$ for $i=1,\ldots,n$. When the model is reflexively plain, this was worked out explicitly in \cite{Melnikov:2010sa}. In terms of the parameters described in section~\ref{ss:param}, the singular locus is given by
\begin{equation}
\label{eq:B2locus}
\prod_{m\neq 0} \left[ \frac{\bsigmah\cdot \bgammah^m}{\bsigmah\cdot\bdeltah}\right]^{\Qh^{\ah}_m} = \kappah_{\ah},
\end{equation}
for some non-zero $\bsigmah$.  Like its A/2-twisted cousin it interpolates between two familiar notions: singularities of $M$ and singularities of $\cF$. On the $(2,2)$ locus, the singularity occurs in the complex structure of $M$; off the $(2,2)$ the singularity may indicate a jump in the rank of the bundle $\cF$. An important difference between the A/2-twisted singular locus and the B/2-twisted singular locus is the latter is determined purely by classical computations while the former includes all the quantum effects.

\subsection{Linear Sigma Model Mirror Map}
\label{ss:mirror}
Mirror symmetry in $(2,2)$ models is the statement that two compactifications on topologically distinct Calabi-Yau's $$M\leftrightarrow M^\circ$$
yield identical four-dimensional field theories. This in turn implies a pairing between Calabi-Yau's---a result of striking mathematical importance. In the heterotic string, one has an additional degree of freedom---a holomorphic vector bundle. A fascinating question that has been generating some interest of late {\em is there a generalisation of mirror symmetry to the heterotic string?} Presumably mirror symmetry would involve a pair-wise exchange of Calabi-Yau together with a choice of vector bundle:\footnote{The fact the mirror exchange would occur pairwise is not obvious. For example, there might a triplet of compactifications giving rise to identical four-dimensional field theories.}
\be
(M, \cF) \leftrightarrow (M^\circ,\cF^\circ)
\ee
For $(0,2)$-models attained as deformations of $(2,2)$ models, the answer to this question is in some sense already known. Mirror symmetry for Calabi-Yau's on the $(2,2)$-locus implies the a pair of identical SCFTs. Consequently, the number of deformations of the $M$ SCFT and the $M^\circ$ SCFT must be identical. However, what is not obvious is whether these deformations are realised by the linear sigma model for each of these Calabi-Yau's. In the $(2,2)$ case we were somewhat lucky: the moduli space probed by the linear sigma model (toric K\"ahler deformations and monomial complex structure deformations) was preserved by the mirror map. Is the same true for bundle deformations?

When the $M$-model is reflexively plain it is argued in \cite{Melnikov:2010sa} the answer to this question is yes. In fact, it is possible to write down an explicit map showing how the linear sigma model parameters map between each other.\footnote{These UV parameters are related to the moduli of the underlying SCFT by a non-trivial RG flow. For $(2,2)$ theories this RG flow is computable via mirror symmetry. We do not yet have the technology to extend this analysis even to reflexively plain models that are deformations of $(2,2)$ theories. } This map on the $(2,2)$ locus was known as the monomial divisor mirror map, and was written down in \cite{Morrison:1995yh}. The generalisation to reflexively plain models is nicely outlined in \cite{Melnikov:2010sa}. The basic idea, as with $(2,2)$ linear sigma models, is that if we define two linear sigma models, the $M$-model and the $M^\circ$-model, these should lead to isomorphic SCFTs. The relation between the parameters defining these two linear sigma models is known as the algebraic mirror map (or the monomial divisor mirror map). Under this map the linear sigma model moduli spaces of the two theories are exchanged, as are the chiral rings. This is reflected by the A/2-twist of the $M$-model being identical to B/2-twist of the $M^\circ$-model. It is not yet known how the chiral rings are exchanged, however it is known how to map the parameters defining the relevant sigma models.

Of course, we are not mapping the full moduli space---instead we are restricting to the subset of parameters that are realised by the linear sigma model. For $(2,2)$ models these are known as toric K\"ahler deformations (K\"ahler deformations of $M$ that descend from the ambient toric variety $V$) and polynomial complex structure deformations (deformations of $M$ realised by coefficients of monomials in $P=0\subset V$). For $(0,2)$ models one is restricted to E-parameters and J-parameters---deformations representable as a monad sequence---and it is these parameters we will map.

In the coordinates of the end of section~\ref{ss:param} the conjectured map takes an elegant form.
\vskip0.1cm
\noindent{\em The mirror map amounts to exchanging $\Delta,\Delta^\circ$, transposing the matrix $b$, and exchanging $\kappa_a$ and $\hat\kappa_{\hat a}$.}

\vskip0.1cm \noindent A first check of this map is that it reduces to the usual monomial divisor mirror map on the $(2,2)$ locus. This is clear once one realises the matrix $b$ reduces to the matrix $\pi_{m i} = \langle m,i\rangle$ defined in \cite{Morrison:1995yh} where $m\in\Delta$ and $i\in\Delta^\circ$. A more non-trivial check is to check that the singular loci of the underlying SCFTs are mapped. That is, the $M$-model A/2-twisted singular locus is exchanged with $M^\circ$-model B/2-twisted singular locus and vice-versa. However, the parameterisation of section~\ref{ss:param} makes this manifest: one sees that \C{eq:A2locus} and \C{eq:B2locus} are exchanged under the map. Thus, the conjectured mirror map looks extremely plausible. However, one still needs to check that the $A/2$-twisted and $B/2$-twisted observables are exchanged under the map, and what the RG flow of the UV parameters is---these remain open questions.

\section{$(0,2)$ Landau-Ginzburg Theories}
\label{s:lg}
One of the fascinating features of $(0,2)$ models is that they exhibit a range of novel effects not present in $(2,2)$ theories. These are most transparent in  $(0,2)$ Landau-Ginzburg theories, where one often has a greater degree of computability and dynamical control. As discussed in section~\ref{ss:phases}, a linear sigma model may exhibit a phase in which the UV dynamics is well-described by a Landau-Ginzburg theory. This is a theory of fields with linear kinetic terms coupled by a superpotential, which in the case of the quintic, takes the form of eq.~\C{eq:lgdef}. In this section we will very briefly summarise some of the developments, old and new, in understanding the structure of $(0,2)$ Landau-Ginzburg theories.

In the early days of $(0,2)$ linear sigma models, there was a lot of interest in Landau-Ginzburg theory, mainly because many properties are exactly computable. For example, one is able to straightforwardly compute the spectrum of states as was done in a series of examples in \cite{Witten:1993yc,Distler:1993mk,Kachru:1993pg}, and later extended by \cite{Distler:1994hq} who, using various symmetries, argue that there are deformations of certain $(2,2)$ Landau-Ginzburg theories that are exactly marginal deformations of the underlying SCFT---an important piece of early evidence in showing that such deformations are not destabilised by worldsheet instantons. The flat directions in question are the $E_6$ singlet directions: bundle, complex structure and K\"ahler deformations that preserve the rank of the bundle, and these are argued to be exactly marginal at finite radius when the parent linear sigma model has a single K\"ahler modulus. More recent work, has appeared in \cite{Aspinwall:2010ve} in which the spectrum singlets corresponding to deformations of $(2,2)$ sigma models was computed in a variety of examples with multidimensional K\"ahler moduli spaces. In some cases the existence of certain states in the spectrum were able to be traced over the K\"ahler moduli space, from small radius (Landau-Ginzburg) to large radius (supergravity). The interest in this arises from the jumping in cohomology groups over the moduli space. For example, in the linear sigma model description of the quintic, the singlet spectrum is known to jump at special values of the complex structure when the theory is at the Landau-Ginzburg point. This jumping is attributable to an enhancement of the gauge symmetries at these special values of the complex structure. More intricate examples have also been computed, which show a more elaborate structure as a function of the moduli. There is jumping of the singlets between small radius and large radius limit. In particular, at small radius there are additional states (as in the quintic) that are not present at large radius. What is the fate of these states? Using mirror symmetry, it is argued in \cite{Aspinwall:2010ve} that these additional states acquire a K\"ahler-dependent mass term, becoming massive away from the Landau-Ginzburg point.

It is also possible to compute correlators in certain toy examples. For example, \cite{Melnikov:2007xi} showed how to compute correlators in  massive Landau-Ginzburg theories attained by deformations of $(2,2)$ theories using simple algebraic techniques. This was extended in \cite{Melnikov:2009nh} to theories without a $(2,2)$-locus. Mathematical aspects of A/2-twisted Landau-Ginzburg models has appeared in \cite{Guffin:2008pi}, while work on developing Landau-Ginzburg theories away from exactly soluble points has appeared in \cite{Kreuzer:2009ha}. Such work is a useful warm-up for studying more general linear model constructions away from the Landau-Ginzburg point.

\subsection{$(0,2)$ Topology Change}
Starting with \cite{Aspinwall:1993nu}, topology change has been well-studied in $(2,2)$ theories, where as one varies the K\"ahler moduli of a compactification, the topology of the target space suddenly changes. Although a rather drastic effect from the point of view of spacetime, one can show using mirror symmetry that on the worldsheet all quantities remain finite and well-defined. This phenomenon may also be present in $(0,2)$ theories, though with a much richer structure. The study of topology changing effects started with \cite{Distler:1996tj,Chiang:1997kt} who analysed what happens to $(0,2)$ theories when $M$ develops certain types of singularities. $(2,2)$ theories on such singular spaces are perfectly well-behaved. Similarly,  \cite{Distler:1996tj,Chiang:1997kt} shows that $(0,2)$ models are also well-behaved. However, unlike in $(2,2)$ theories there may be multiple resolutions of the singularities. Essentially, one needs to lift the bundle on $M$ to the resolved space, and there can be multiple ways of doing this. Although these resolutions may look distinct from the point of view of spacetime, they are described by the same Landau-Ginzburg theory. The interpretation proposed by \cite{Distler:1996tj,Chiang:1997kt} is that the Landau-Ginzburg point is where a perturbative string transition takes place (i.e. topology change). However, later analysis of these Landau-Ginzburg theories by  \cite{Blumenhagen:1997vt,Blumenhagen:1997cn} seemed to indicate the correct interpretation is of a type of string duality taking place, reminiscent of mirror symmetry. For example, although the compactifications look different at large radius, their moduli spaces are in fact identical. Nonetheless, it still seems likely that topology changing transitions occur in different contexts in $(0,2)$ models, similar to that discussed in the $(2,2)$ context in \cite{Aspinwall:1993nu}, in which there is an enlarged K\"ahler moduli space, with walls separating different domains. Clarifying precisely if and how this happens, as well as what types of topology changes occur in generic $(0,2)$ models remains an open question.

\subsection{Mirror Pairs}
As we discussed above mirror symmetry in the setting of heterotic string compactifications is a fascinating, old and yet open question. From the point of view of the linear sigma model, \cite{Melnikov:2010sa} constructed a simple map for reflexively plain models attained as deformations of $(2,2)$ sigma models. What about the more general setting? What if the models are not reflexively plain? What if there is not even a $(2,2)$ locus?

Some progress has been made in this direction in the context of $(0,2)$ Landau-Ginzburg theories. It was first discussed in the context of $(0,2)$ Landau-Ginzburg theories by \cite{Distler:1995bc}. The situation described there involved studying the GLSM for two particular topologically distinct target spaces. On a first glance it appears the GLSMs are quite different for the two different compactifications. However, if one goes to the Landau-Ginzburg point, $r=-\infty$, then the models become identical and are related by a simple map. The map involved an exchange of complex structure moduli with moduli corresponding to deforming the holomorphic structure of the bundle. One then argues that the theories have identical perturbative expansions in an open set around the Landau-Ginzburg point $r=-\infty$, and consequently agree on the whole $r,\theta$ plane (see also \cite{Blumenhagen:1997vt,Blumenhagen:1997cn} for later discussions). Mirror symmetry in $(0,2)$ Landau-Ginzburg theories has been followed up by numerous papers. Using a class of $(0,2)$ Landau-Ginzburg theories developed in \cite{Blumenhagen:1995ew,Blumenhagen:1996xb}, \cite{Blumenhagen:1996vu} proposed a version of mirror symmetry which exchanged complex structure, K\"ahler and bundle moduli. In \cite{Blumenhagen:1997pp,Blumenhagen:1996tv} additional examples of Landau-Ginzburg mirror pairs were constructed using orbifolding techniques similar to that of Greene-Plesser\cite{Greene:1990ud}, while some recent discussion of Landau-Ginzburg models in relation to mirror symmetry has appeared in \cite{Kreuzer:2009ha} and a computation of elliptic genera in $(0,2)$ Landau-Ginzburg theories has recently appeared in \cite{Ando:2009av}.

One drawback of most of the discussion of $(0,2)$ mirror symmetry to date, is that it is limited to Landau-Ginzburg theories, a very special type of a string compactification. Aside from the map in discussed in section~\ref{ss:mirror}, there are a couple of other notable exceptions. One proposal for extending the monomial-divisor mirror map to $(0,2)$ theories has appeared in \cite{Sharpe:1998wa}. In \cite{Adams:2003zy}, a proposal for extending the abelian duality of \cite{Hori:2000kt,Morrison:1995yh} to $(0,2)$ theories was developed. This duality is a relation between massive $(0,2)$ GLSMs and Landau-Ginzburg theories. It exchanges charged fields with uncharged fields, and is generated by dualising the $\U(1)$ torus action of the toric variety. As the torus action is not free, a non-perturbative superpotential is generated. In some sense this is to be thought of as mirror symmetry for massive models. The proposal of \cite{Adams:2003zy} was checked in \cite{Katz:2004nn,Guffin:2007mp} by computing worldsheet instanton corrections to correlators in the corresponding $(0,2)$ sigma model. To better understand mirror symmetry in more general $(0,2)$-models one really needs a generalisation of the $(2,2)$ algebraic map of \cite{Morrison:1995yh,Morrison:1994fr} to include $(0,2)$ moduli, followed by a calculation of the spacetime D-terms, which would determine the renormalisation of the algebraic parameters $q_a$ of the sigma model, to the generalisation of the ``special coordinates'' of the SCFT.

\section{Outlook}
\label{s:outlook}
In this review we have touched on some recent work aimed at uncovering the worldsheet structure of $(0,2)$ heterotic worldsheet theories. $(0,2)$ theories have undergone a remarkable cycle of dying and then rising from the dead. This means there are many open questions to be tackled in this field. Indeed, the past couple of years have witnessed a revival, with an increasing amount of work aimed at uncovering their hidden gems.

What are some of the remaining open issues? One is that of $(0,2)$ mirror symmetry. What is the form of mirror symmetry for non reflexively plain models? What about if there isn't a $(2,2)$ locus, what form does mirror symmetry take then? Another is that even supposing we understand (0,2) mirror symmetry for linear sigma models, to make contact with physical observables, we will still have match the linear model deformations to moduli of the SCFT, and determine the K\"ahler potential.  These are not easy tasks, but our success gives us hope that perhaps even in questions regarding the K\"ahler potential progress may be made by considering additional structure beyond (0,2) supersymmetry in these vacua.  Perhaps these additional structures (such as the $\GU(1)_L$ current algebra) may enable us to extend some of the results of~\cite{Dixon:1989fj} off the (2,2) locus. This would be important phenomenologically, as we would then have computed normalised Yukawa couplings, and taken a step closer towards connecting the heterotic worldsheet with its supergravity counterpart.

\newpage
\providecommand{\href}[2]{#2}\begingroup\raggedright\endgroup
%\bibliographystyle{utphys}
%\bibliography{review}
\end{document}